\documentclass[twocolumn,aps,prd,amssymb,eqsecnum,nofootinbib,superscriptaddress,showpacs]{revtex4}
\usepackage{hyperref}

\usepackage{amsmath}
\usepackage{amssymb}
\usepackage{epsfig}
\usepackage{bm}
\usepackage{slashbox}
\usepackage[usenames,dvipsnames]{color}
\usepackage{graphicx}
\usepackage{epsfig}
\usepackage{psfrag}
\usepackage{rotating}
\usepackage{multirow}

\newcommand{\bma}{\begin{pmatrix}}
\newcommand{\ema}{\end{pmatrix}}
\newcommand{\bea}{\begin{eqnarray*}}
\newcommand{\eea}{\end{eqnarray*}}

\newcommand{\Caltech}{\affiliation{Theoretical Astrophysics 350-17,
    California Institute of Technology, Pasadena, California 91125, USA}}
\newcommand{\Cornell}{\affiliation{Center for Radiophysics and Space
    Research, Cornell University, Ithaca, New York 14853, USA}}
\newcommand{\NITHEP}{\affiliation{National Institute of Theoretical Physics, Private Bag X1 Matieland, Stellenbosch 7602, South Africa}}
\newcommand{\Stias} {\affiliation{
Stellenbosch Institute for Advanced Study (STIAS),
Wallenberg Research Centre at Stellenbosch University,
Marais Street,
Stellenbosch 7600,
South Africa}}

\begin{document}

\title{Visualizing Spacetime Curvature via Frame-Drag Vortexes and Tidal Tendexes \\ I. General Theory and Weak-Gravity Applications}

\author{David A.\ Nichols} \Caltech
\author{Robert Owen} \Cornell
\author{Fan Zhang} \Caltech
\author{Aaron Zimmerman} \Caltech
\author{Jeandrew Brink} \NITHEP
\author{Yanbei Chen} \Caltech
\author{Jeffrey D.\ Kaplan} \Caltech
\author{Geoffrey Lovelace} \Cornell
\author{Keith D.\ Matthews} \Caltech
\author{Mark A.\ Scheel} \Caltech
\author{Kip S.\ Thorne} \Caltech \Stias
\date{printed \today}

\begin{abstract}

When one splits spacetime into space plus time, 
the Weyl curvature tensor (vacuum Riemann tensor) gets 
split into two spatial, symmetric, and trace-free (STF) tensors:
(i) the Weyl tensor's so-called ``electric'' part or {\it tidal field} 
$\mathcal E_{jk}$, which raises tides on the Earth's
oceans and drives geodesic deviation (the relative acceleration of
two freely falling test particles separated by a spatial vector $\xi^k$ is
$\Delta a_j = -\mathcal E_{jk} \xi^k$);  
and (ii) the Weyl tensor's so-called ``magnetic'' part or (as we call it)
{\it frame-drag field} $\mathcal B_{jk}$, which 
drives differential frame dragging
(the precessional angular velocity of a gyroscope at the tip of $\xi^k$,
as measured using a local inertial frame at the tail of $\xi^k$, is 
$\Delta \Omega_j = \mathcal B_{jk} \xi^k$).  

Being STF, $\mathcal E_{jk}$ and $\mathcal B_{jk}$ each have three orthogonal
eigenvector fields which can be depicted by their integral curves.  We call the integral curves
of $\mathcal E_{jk}$'s eigenvectors {\it tidal tendex lines} or simply {\it tendex lines}, we call each tendex line's eigenvalue its {\it tendicity}, and we 
give the name {\it tendex} to a collection of tendex lines
with large tendicity.
The analogous quantities for $\mathcal B_{jk}$ are 
{\it frame-drag vortex  lines} or simply {\it vortex lines}, their {\it vorticities}, and 
{\it vortexes}.   

These concepts are powerful tools for  visualizing spacetime curvature.
We build up physical intuition into them 
by applying them to a variety
of weak-gravity phenomena:  a spinning, gravitating point particle, two such particles
side-by-side, a plane gravitational wave, a point particle with 
a dynamical current-quadrupole moment or dynamical 
mass-quadrupole moment, and a slow-motion binary system made of 
nonspinning point
particles.  We show that a rotating current quadrupole has four
rotating vortexes that sweep outward and backward like water streams from 
a rotating sprinkler.
As they sweep, the vortexes acquire 
accompanying tendexes and thereby become outgoing 
current-quadrupole gravitational waves.
We show similarly that a rotating mass quadrupole has four rotating, 
outward-and-backward sweeping 
tendexes that acquire accompanying vortexes as they sweep, and become outgoing
mass-quadrupole gravitational waves.  We show, further,  that an oscillating current quadrupole
ejects sequences of vortex loops that acquire accompanying tendex loops 
as they travel, and become current-quadrupole gravitational waves;  
and similarly for an oscillating mass quadrupole. And we show how a binary's
tendex lines transition, as one moves radially, from those of two 
static point particles in the deep near zone, to those of a single 
spherical body in the outer part of the near zone and inner part of the wave
zone (where the binary's mass monopole moment dominates), to those of a
rotating quadrupole in the far wave zone (where the quadrupolar
gravitational waves dominate). 

In paper II we will use these vortex and tendex concepts to gain insight 
into the quasinormal modes of
black holes, and in subsequent papers, by combining these concepts with 
numerical
simulations, we will explore the nonlinear 
dynamics of curved spacetime around colliding black holes.  We have published a
brief overview of these applications in {\it Physical Review Letters}~\cite{OwenEtAl:2011}. We expect these vortex and tendex concepts to become powerful tools for general relativity research in a variety of topics.  

\end{abstract}

\pacs{04.25.dg, 04.25.Nx, 04.30.-w}

\maketitle

\section{Motivation and Overview}
\label{sec:intro}

In the 1950s John Archibald Wheeler coined the phrase {\it geometrodynamics} to 
epitomize his intuition that curved spacetime must have a rich range of nonlinear dynamical
behaviors --- behaviors that are important in our Universe and are worthy of probing deeply by both
theoretical and observational means (see Ref.~\cite{Geometrodynamics} and
earlier papers by Wheeler reprinted therein and also 
Ref.~\cite{WheelerLesHouches63}). 
It was obvious to Wheeler that analytical tools by themselves would not be sufficient
to reveal the richness of geometrodynamics, so he encouraged his
colleagues and students to begin developing 
numerical tools~\cite{misner1960,BrillLindquist1963,HahnLindquist1964}, and he encouraged
Joseph Weber to develop technology for gravitational-wave 
observations~\cite{Weber1959}. 

Today, a half century later, numerical relativity has finally reached 
sufficient
maturity (for a review, see Ref.~\cite{Centrella:2010} 
and the references therein) that, 
hand in hand with analytical relativity, 
it can be used to explore
nonlinear geometrodynamics in generic situations; and
gravitational-wave detectors are
sufficiently 
mature~\cite{Barish:1999,Sigg:2008,Acernese:2008,Kuroda:2010,Harry2010} 
that they may soon observe nonlinear 
geometrodynamics in black-hole collisions.

Unfortunately, there is a serious obstacle to extracting geometrodynamical insights
from numerical-relativity simulations: a paucity of good tools for visualizing 
the dynamics of curved spacetime.  We are reasonably sure that buried in the billions of numbers
produced by numerical-relativity simulations there are major discoveries to be
made, but extracting those discoveries is exceedingly difficult and perhaps impossible
with the tools we have had thus far.

Until now, curved spacetime has been visualized primarily via 
(isometric) {\it embedding diagrams}~(Sec. 23.8 of Ref.~\cite{MTW}):
choosing spacelike two-dimensional surfaces in spacetime, 
and embedding them in flat 3-dimensional Euclidean space or 
2+1-dimensional Minkowski spacetime
in a manner that preserves the 
surfaces' intrinsic geometry. 
(For some examples of embedding diagrams applied 
to black-hole spacetimes, see, 
e.g., Refs.~\cite{Smarr:1973,Marolf1999,JacobPiran2006}).
Unfortunately, such embedding diagrams are of very limited value.  
They capture
only two dimensions of spacetime, and the 2-surfaces of greatest interest
often cannot be embedded globally in flat Euclidean 3-space or 
flat Minkowski 
2+1-dimensional spacetime~\cite{Smarr:1973,RomanoPrice1995,
Mihai2002,Chan2006}. 
Mixed Euclidean/Minkowski embeddings are often required 
(e.g., Fig.~4 of Ref.~\cite{Smarr:1973}), and such embeddings have not
proved to be easily comprehended. 
Moreover, although it is always possible to perform a local embedding
in a flat 3-space (in the vicinity of any point on the two-surface), when one 
tries to extend the embedding to cover the entire two-surface, one often 
encounters discontinuities analogous to shocks in fluid mechanics 
\cite{RomanoPrice1995,Chan2006}.

A systematic approach to understanding the connection between nonlinear 
near-field dynamics in general relativity and emitted gravitational waves
is being developed by Rezzolla, Jaramillo, Macedo, and 
Moesta~\cite{Rezzolla2010,Jaramillo2011a,Jaramillo2011b,Jaramillo2011c}.  
This approach focuses on correlations between data on a surface at large 
radius (ideally null infinity) and data 
on world tubes in the source region (such as black-hole horizons).  
The purpose is to use such correlations to 
infer the dynamics of a black hole (e.g. the kick) directly 
from data on its horizon.  While we find this approach exciting and 
attractive, in our own work we seek a more direct set of tools: 
tools that can probe the dynamics of spacetime curvature that cause 
such correlations in the first place, and that can be more readily and 
intuitively 
applied to a wider range of other geometrodynamic phenomena.  
It is our hope that eventually our tools and those of  
Rezzolla et.~al.~\cite{Rezzolla2010,Jaramillo2011a,Jaramillo2011b} 
will provide complementary pictures for understanding spacetime dynamics, 
and particularly black-hole kicks.

We have introduced our new set of tools in
a recent paper in {\it Physical Review Letters} \cite{OwenEtAl:2011}.  They 
are tools for visualizing spacetime curvature, called
{\it tidal tendex lines, tendicities, and tendexes}; and {\it frame-drag vortex lines, vorticities
and vortexes}. These tools capture the full details of the Weyl curvature
tensor (vacuum Riemann tensor), which embodies spacetime curvature. They do
so in three-dimensional, dynamically evolving pictures, of which snapshots can be 
printed in a paper such as this one, and movies can be made available 
online.\footnote{Just as there is no unique method to evolve field lines
in electromagnetism, so too is there no unique way to match tendex or vortex
lines at one time with others at a later time.
Nevertheless, animations of field lines are useful for pedagogical purposes 
and for building intuition \cite{Belcher2003}.
While some of the authors and colleagues are investigating how to evolve
tendex and vortex lines in generic situations, the animations of the 
lines posted online all have special symmetries that provide a natural
way to connect lines at one time with lines at the next.}  
Specifically, as of this writing two movies can be seen at 
Refs.~\cite{RotatingCurrentQuadMovie, AntiAlignedSpinsMovie};
one shows the vortex lines from a rotating current quadrupole, the other, 
vortex lines from two particles that collide head-on with transverse, 
antiparallel spins.

We have found these tools to be an extremely powerful way to visualize the output of numerical simulations. 
We have also used them to obtain deep new insights into old analytical 
spacetimes.  
We have applied them, thus far, to pedagogical linear-gravity problems 
(this paper and \cite{Zimmerman2011}), 
to stationary and perturbed black holes (Paper II in this series), and
to simulations of the inspiral and mergers of spinning 
black holes (\cite{OwenEtAl:2011} and Paper III).
We plan to apply them in the future
in a variety of other geometrodynamical venues, such as black holes ripping apart
neutron stars and curved spacetime near various types of singularities.

This is the first of a series of papers in which we will (i) present these tools, (ii) show
how to use them, (iii) build up physical intuition into them, and (iv) employ them
to extract geometrodynamical insights from numerical-relativity simulations. Specifically: 

In this paper (Paper I), we introduce these vortex and tendex tools, and we then apply them 
to weak-gravity situations (linearized general relativity) with special focus on the roles of 
vortexes and tendexes in gravitational-wave generation. 
In a closely related paper \cite{Zimmerman2011}, three of us have applied these tools
to visualize asymptotic gravitational radiation and explore the topology of its vortex and tendex lines,
and also to explore a linearized-gravity model of an extreme-kick merger.
In Paper II we shall apply our new tools 
to quiescent black holes and quasinormal modes of black holes, with special focus
once again on the roles of vortexes and tendexes in generating gravitational waves.
In Paper III and subsequent papers we shall apply our tools to numerical simulations of 
binary black holes, focusing on nonlinear geometrodynamics in the holes' near zone and
how the near-zone vortexes and tendexes generate gravitational waves.

The remainder of this paper is organized as follows: 

In Sec.\ \ref{sec:3+1Split} we review the well-known split of the Weyl curvature tensor into its 
``electric'' and ``magnetic'' parts $\mathcal E_{ij}$ and $\mathcal B_{ij}$, and
in Sec.\ \ref{sec:MaxwellLikeEqns} we review the Maxwell-like evolution equations for 
$\mathcal E_{ij}$ and $\mathcal B_{ij}$ and discuss the 
mathematical duality between these fields.  Then in Sec.\ \ref{sec:InterpretEandB} we review the well-known
physical interpretation of $\mathcal E_{ij}$ as the 
{\it tidal field} that 
drives geodesic deviation and the not so well-known interpretation of
$\mathcal B_{ij}$ \cite{Estabrook1964,Schmid2009} as the {\it frame-drag}
field that drives differential frame dragging, 
and we derive the equation 
of differential frame dragging. 

In Sec.\ \ref{sec:NewTools} we introduce our new set of tools for visualizing spacetime curvature.
Specifically:  In Sec.\ \ref{sec:Lines} we introduce tendex lines and their tendicities, and we
quantify them by their stretching or compressional force on a person; 
and we also introduce vortex lines and their vorticities and quantify them
by their twisting (precessional) force on gyroscopes attached to the head and feet of
a person.  Then in Sec.\ \ref{sec:VortexesTendexes} we introduce {\it vortexes} and {\it tendexes} 
(bundles of vortex and tendex lines that have large vorticity and tendicity)
and give examples.  

In the remainder of this paper we illustrate these new concepts by applying them
to some well-known, weak-gravity, analytic examples of spacetime curvature.
In Sec.\ \ref{WeakGravityStationary} we focus on the spacetime curvature of stationary systems, and in Sec.\ \ref{sec:GWandGeneration} we focus 
on dynamical systems and develop physical pictures of how they generate 
gravitational waves.

More specifically, in Sec.\  \ref{sec:OneWeakGravityBody}, we compute $\mathcal E_{ij}$ and $\mathcal B_{ij}$
for a static, gravitating, spinning point particle; we explain the relationship of
$\mathcal B_{ij}$ to the particle's dipolar ``gravitomagnetic field,'' we draw
the particle's tendex lines and vortex lines, and we identify two vortexes that
emerge from the particle, a counterclockwise vortex in its ``north polar'' region
and a clockwise vortex in its ``south polar'' region.  In Sec.\ \ref{TwoWeakGravityBodies}, we
draw the vortex lines for
two spinning point particles that sit side-by-side with their spins in opposite directions,
and we identify their four vortexes.  
Far from these particles, they look like a single point particle with a 
current-quadrupole
moment.  In Sec.\ \ref{sec:StationaryCurrentQuadrupole}, we  draw the  vortex lines for
such a current-quadrupole particle and identify their vortexes.  Then in Sec.\ \ref{sec:StaticMassQuadrupole}, we show that the tendex lines of
a mass-quadrupole particle have precisely the same form as the vortex lines of the
current-quadrupole particle, and we identify the mass quadrupole's four tendexes.  

Turning to dynamical situations, in Sec.\ \ref{sec:PlaneWave} we compute 
$\mathcal E_{ij}$ and $\mathcal B_{ij}$ for a plane gravitational wave, 
we express them in terms of the Weyl scalar $\Psi_4$, and we draw their vortex 
and tendex lines.  
In Sec.\ \ref{sec:GWsHeadOnCollision} we explore the quadrupolar 
($l=2$, $m=0$) angular pattern of gravitational waves from the head-on 
collision of two black holes, and we draw their vortex lines and tendex lines, 
intensity-coded by vorticity and tendicity, on a sphere in the wave zone.
In Sec.\ \ref{sec:CurrentQuadrupoleWaveGeneration} we compute $\mathcal E_{ij}$ and $\mathcal B_{ij}$ for
a general, time-varying current-quadrupolar particle, and then in Secs.\ \ref{sec:RotatingCurrentQuadrupole} and \ref{sec:OscillatingCurrentQuadrupole} we
specialize to a rotating current quadrupole and an oscillating current quadrupole, and 
draw their vortex and tendex lines.  Our drawings and the mathematics reveal that
the particle's outgoing gravitational waves are generated by its near-zone vortexes. 
The rotating current quadrupole has four vortexes that spiral outward and backward
like four water streams from a rotating sprinkler.  As it bends backward, each vortex
acquires an accompanying tendex, and the vortex and tendex together become a 
gravitational-wave crest or gravitational-wave trough.  The oscillating current quadrupole,
by contrast, ejects vortex loops that 
travel outward, acquiring accompanying 
tendex loops with strong tendicity on the transverse segment of each loop
and weak on the radial segment---thereby becoming outgoing gravitational waves.

In Sec.\ \ref{sec:MassQuadrupoleWaveGeneration}
we show that a time-varying mass quadrupole produces the same phenomena as
a time-varying current quadrupole, but with vortexes and tendexes interchanged.

In Sec.\ \ref{sec:Binary} we study the vortexes and tendexes of a slow-motion
binary made of nonspinning point particles.  In the near zone, the tendex lines
transition, as one moves radially outward, from those of two 
individual particles 
(radial and circular lines centered on each particle) toward those of a single
spherical body (radial and circular lines centered on the binary and produced
by the binary's mass monopole moment).  In the transition zone and inner
wave zone, the mass monopole continues to dominate. Then at radii $r\sim a^2/M$
(where $a$ is the particles' separation and $M$ is the binary's mass), the
radiative quadrupole moment begins to take over and the tendex lines
gradually transition into the outward-and-backward spiraling lines of a
rotating quadrupole.  

We make some concluding remarks in Sec.\ \ref{sec:Conclusions}.

Throughout this paper we use geometrized units with $c=G=1$, and we use the 
sign conventions of MTW~\cite{MTW} for the metric signature, the Weyl 
curvature, and the Levi-Civita tensor. 
We use Greek letters for spacetime indices (0--3) and Latin letters for 
spatial indices (1--3), and we use arrows over 4-vectors and bold-face font for
spatial 3-vectors and for tensors.  In orthonormal bases, we use hats 
over all kinds of indices. 

\section{The tidal field $\mathcal E_{ij}$ and frame-drag field $\mathcal B_{ij}$}
\label{sec:EandB}

\subsection{3+1 split of Weyl curvature tensor into \\
$\mathcal E_{ij}$ and $\mathcal B_{ij}$}
\label{sec:3+1Split}

For a given spacetime, the Weyl curvature tensor can be calculated
from the Riemann tensor by subtracting Riemann's trace from itself;
i.e., by subtracting from Riemann the following
combinations of the Ricci curvature
tensor $R^{\mu}_{\phantom{a}\nu}$, and Ricci curvature scalar
$R$ (Eq.~(13.50) of MTW~\cite{MTW}):
\begin{equation}  \label{eq:DefWeyl}
C^{\mu \nu}_{\phantom{ab}\rho \sigma} = R^{\mu \nu}_{\phantom{ab}\rho \sigma} -2 \delta^{[
   \mu}_{\phantom{ab}[ \rho}R^{\nu]}_{\phantom{ij} \sigma]} +
\frac{1}{3} \delta^{[ \mu}_{\phantom{ab}[ \rho}\delta^{\nu
  ]}_{\phantom{ab}\sigma ]}R\;.
\end{equation}
Here $\delta^\mu_{\phantom{a}\rho}$ is the Kronecker delta, and the square
brackets represent antisymmetrization. Note that in
vacuum, $C^{\mu \nu}_{\phantom{ab}\rho\sigma} = R^{\mu
  \nu}_{\phantom{ab}\rho \sigma}$, and thus in vacuum the Weyl
tensor contains all information about the spacetime curvature.  

Let us pick a foliation of spacetime into a family of spacelike
hypersurfaces.  We shall denote by 
$u^\mu$ the 4-velocity of observers who move orthogonal to the foliation's 
space slices, and by $\gamma_{\mu\nu} = g_{\mu\nu} + u_\mu u_\nu$ the 
induced spatial three metric on these slices, so that
$\gamma_\alpha^{\phantom{a} \mu}$ is the projection operator onto the slices.  
As is well-known, e.g.~\cite{Maartens1998},
using this projection operator,
we can split the Weyl tensor
covariantly into two irreducible parts, which are
symmetric, trace-free (STF) tensors that lie in the
foliation's hypersurfaces (i.e.\ that are orthogonal to $u^\mu$). 
These pieces are
\begin{subequations}
\label{eq:DefsEijBij}
\begin{equation}  \label{eq:DefEij}
\mathcal E_{\alpha\beta} = {\gamma_\alpha}^\rho {\gamma_\beta}^\sigma 
C_{\rho\mu\sigma\nu}u^\mu u^\nu\;, \quad \hbox{i.e. }
\mathcal E_{ij} = C_{i \hat 0 j \hat 0}\;, 
\end{equation}
an even-parity field called the ``electric'' part of
$C^{\mu \nu}_{\phantom{ab}\rho\sigma}$, and
\begin{equation}  \label{eq:DefBij}
\mathcal B_{\alpha\beta} = - {\gamma_\alpha}^\rho {\gamma_\beta}^\sigma
\,^*C_{\rho\mu\sigma\nu}u^\mu u^\nu\;, \quad \hbox{i.e. }
 \mathcal B_{ij} = \frac{1}{2} \epsilon_{ipq}C^{pq}_{\phantom{ab}j\hat 0}, 
\end{equation}
\end{subequations}
an odd-parity field known as the ``magnetic'' part of
$C^{\mu \nu}_{\phantom{ab}\rho \sigma}$.  Here
the symbol * represents the (left) Hodge dual, 
$^*C_{\rho\mu\sigma\nu} = 
\frac12 \epsilon_{\rho\mu\eta\lambda} {C^{\eta\lambda}}_{\sigma\nu}$,
and for each field the second expression is written in 3+1 notation:
the Latin (spatial)
indices are components in the foliation's hypersurface, and the
$\hat 0$ is a component on the foliation's unit time basis vector 
$\vec e_{\hat 0} \equiv \vec u$. Our normalization for the Levi-Civita tensor is that of MTW: 
in a right-handed orthonormal frame,  $\epsilon_{\hat 0 \hat 1 \hat 2 \hat 3}= +1$, 
and the spatial Levi-Civita tensor is defined by $\epsilon_{ipq} = \epsilon_{\hat 0 ipq}$, 
with $\epsilon_{\hat 1 \hat 2 \hat 3} = 1$ in a right-handed orthonormal basis. 
Note that Eqs.\ (\ref{eq:DefsEijBij}) are a direct and intentional analogy to the decomposition 
of the Maxwell tensor of
electromagnetism $F_{\mu \nu}$ into the familiar electric and
magnetic fields $E_i$ and $B_i$ \cite{Maartens1998}:
\begin{equation}  \label{eq:DefEiBi}
E_i = F_{i\hat 0}\;, \;\;\;\;  B_i = -\,^*F_{i\hat 0} = \frac{1}{2}\epsilon_{ipq} F^{pq}.
\end{equation}
Note that our sign conventions differ from \cite{Maartens1998}, where $\epsilon_{\hat 0 \hat 1 \hat 2 \hat 3} = -1$, and so Eq.~\eqref{eq:DefBij} has an additional minus sign in order to maintain a strict analogy with the magnetic field $B_i$ of electromagnetism. This results in a 
$\mathcal{B}_{ij}$ defined with a different sign convention than, for example, in \cite{Penrose1992,Stephani2003}.

\subsection{Evolution of $\mathcal E_{ij}$ and $\mathcal B_{ij}$}
\label{sec:MaxwellLikeEqns}

The propagation equations for the Weyl tensor and its 
gravito-electromagnetic 
representation are the Bianchi identities.  We shall write them
down and discuss them in three contexts: a general foliation and
coordinate system, 
the local-Lorentz frame of
a freely falling observer, 
and the weak-gravity, nearly Minkowski
spacetimes of the current paper (Paper I in this series).

\subsubsection{General foliation and coordinate system in the language of numerical relativity}

Because this paper is a foundation for using $\mathcal E_{ij}$ and
$\mathcal B_{ij}$ to interpret the results of numerical-relativity simulations,
we shall write their evolution equations (the Bianchi identities)
in a general coordinate
system of the type used in numerical relativity, and we shall discuss
these equations' mathematical structure in the language of numerical
relativity.

We denote by $t$ a time coordinate that
is constant on the foliation's hypersurfaces, and by $\alpha$ and $\vec\beta$
the foliation's lapse and shift functions, so the 
orthogonal observers' 4-velocity is 
${\vec u} = \alpha^{-1} ({\vec \partial}_t - \vec \beta)$. 
The 3+1 split divides the Bianchi identities into evolution equations that govern the time evolution of the spatial fields, and constraint equations that are obeyed by the fields on each time slice. The evolution equations are~\cite{Friedrich96,Anderson98} 
\begin{equation} \label{eq:Bianchi31}
\begin{split}   
\partial_t \mathcal{E}_{ij} =& \mathcal{L}_{\beta} \mathcal{E}_{ij} 
                 + \alpha [D_k \mathcal{B}_{l(i}\epsilon_{j)}^{\phantom{i}kl} - 3 \mathcal{E}^k{}_{(i}K_{j)k} \\
                 &+ K^k{}_k \mathcal{E}_{ij}  - \epsilon_i^{\phantom{i}kl} \mathcal{E}_{km} K_{ln} \epsilon_j^{\phantom{i}mn}
                 + 2 a_k \mathcal{B}_{l(i}\epsilon_{j)}^{\phantom{i}kl}] \,, \\
\partial_t \mathcal{B}_{ij} =& \mathcal{L}_{\beta} \mathcal{B}_{ij} 
                + \alpha  [- D_k \mathcal{E}_{l(i}\epsilon_{j)}^{\phantom{i}kl} - 3 \mathcal{B}^k{}_{(i}K_{j)k} \\
                & + K^k{}_k \mathcal{B}_{ij} 
                - \epsilon_i^{\phantom{i}kl} \mathcal{B}_{km} K_{ln} \epsilon_j^{\phantom{i}mn}
                - 2 a_k \mathcal{E}_{l(i}\epsilon_{j)}^{\phantom{i}kl}] \,.
\end{split}
\end{equation}
Here the extrinsic curvature, Lie derivative on a second rank tensor, and acceleration of the slicing are respectively defined by
\begin{eqnarray}
\label{eq:3+1Defs}
K_{i j} & = & -\frac{1}{2\alpha}(\partial_t \gamma_{ij} - D_i \beta_j - D_j \beta_i) \,, \\
\mathcal{L}_{\beta} \mathcal E_{ij}& = & \beta^k D_k \mathcal E_{ij} + \mathcal E_{ik} D_j \beta^k +   \mathcal E_{kj} D_i \beta^k \,, \\
a_k & =&  D_k \ln \alpha \,.
\end{eqnarray}
The derivative $D_i$ is the covariant derivative  associated with the induced metric $\gamma_{ij}$ on the slices. 
The evolution system~(\ref{eq:Bianchi31}) is closed by an additional
evolution equation for the 3-metric, which is Eq.~(\ref{eq:3+1Defs}), and 
evolution equations for the extrinsic curvature and 
the 3-dimensional connection $\Gamma^k_{ij}$, which are

\begin{equation}
\begin{split}
\partial_t K_{ij} =& \mathcal{L}_{\beta} K_{ij}
-\alpha[ \partial_k \Gamma^k_{ij} 
  -\Gamma^k_{lj}\Gamma^l_{ki}
  +\partial_i\partial_j q 
  \\ &
  +\partial_i \ln \alpha \, \partial_j \ln \alpha
  -\Gamma^k_{ij}\partial_k q 
  -2 \mathcal{E}_{ij} 
  +K^k{}_k K_{ij} 
  ], \\
\partial_t \Gamma^k_{ij} =&
\mathcal{L}_{\beta} \Gamma^k_{ij}
- \alpha D^k K_{ij} + K_{ij} D^k \alpha- 2 K^k{}_{(i}D_{j)}\alpha
  \\ &
+2\alpha \epsilon^{kl}{}_{(i}\mathcal{B}_{j)l},
\end{split}
\end{equation}
where we have defined
\begin{eqnarray}
  q &=& \ln(\alpha \gamma^{-1/2}),\\
  \mathcal{L}_{\beta} \Gamma^k_{ij} &=& 
  \beta^l\partial_l \Gamma^k_{ij}+2\Gamma^k_{l(j}\partial_{i)}\beta^l
  \nonumber \\ &&
  -\Gamma^l_{ij}\partial_l \beta^k + \partial_i\partial_j\beta^k.
\end{eqnarray} 
The above equations are symmetric hyperbolic if
$q$ and $\beta^i$ are specified functions of time and space.

The constraint equations on each slice are the definitions of 
$\mathcal{E}_{ij}$ and $\mathcal{B}_{ij}$,
\begin{equation}
\label{eq:EinsteinConstraints}
\begin{split}
\mathcal{E}_{ij} &= {}^{(3)}R_{ij} + K^k{}_k K_{ij} - K^k{}_i K_{jk},\\
\mathcal{B}_{ij} &= \epsilon_j{}^{lk}D_k K_{li},
\end{split}
\end{equation}
from which the Einstein constraints follow from the condition that 
$\mathcal{E}_{ij}$ and $\mathcal{B}_{ij}$ are symmetric and trace-free,
and the definition of $\Gamma^k_{ij}$,
\begin{equation}
  \label{eq:ChristoffelDefinition}
  \Gamma^k_{ij} = \frac{1}{2} \gamma^{k\ell}\left(\partial_i \gamma_{j\ell}
  +\partial_j \gamma_{i\ell} - \partial_\ell\gamma_{ij}\right).
\end{equation}
The Bianchi identities imply derivative
constraints on $\mathcal{E}_{ij}$ and $\mathcal{B}_{ij}$:
\begin{equation}
 \label{eq:BianchiConstraint} 
\begin{split}   
D^i \mathcal{E}_{ij} &= \mathcal{B}_{ik}K^i{}_l \epsilon^{kl}{}_j \,, \\ 
D^i \mathcal{B}_{ij} &= - \mathcal{E}_{ik}K^i{}_l \epsilon^{kl}{}_j \,.
\end{split}
\end{equation}   
These last equations are automatically satisfied
if Eqs.~(\ref{eq:EinsteinConstraints}) are satisfied. 
Equations.~(\ref{eq:BianchiConstraint}) are nonlinear, but otherwise they have the same structure as the constraints in simple electromagnetism.

Note also that the equations governing $\bm{\mathcal E}$ and $\bm{\mathcal B}$, Eqs~(\ref{eq:Bianchi31}) and (\ref{eq:BianchiConstraint}) share another similarity with the field equations of electromagnetism. Namely, just as the Maxwell equations are invariant under the duality transformation
\bea
{\bm E} \rightarrow {\bm B}\;, \quad \quad {\bm B} \rightarrow -{\bm E}\;,
\eea
i.e.\ under a rotation in the complexified notation 
\bea
{\bm E} - i {\bm B} \rightarrow e^{i\pi/2}({\bm E} - i {\bm B})\;,
\eea
so the exact Maxwell-like Bianchi identities 
(\ref{eq:Bianchi31}) are also invariant under the same duality transformation 
\begin{equation} \label{eq:DualityRotation}
\bm {\mathcal{E} } \rightarrow \bm{\mathcal{B}}\;, \quad
\bm{\mathcal{B}} \rightarrow - \bm{ \mathcal{E}}\;. 
\end{equation}

This {\it duality} in the structure of Eqs.~(\ref{eq:Bianchi31})
and also~(\ref{eq:BianchiConstraint}) does not in general enable one to
construct one metric solution of Einstein's equations from another, known
solution.  However, as we shall see, we can utilize this duality in weakly gravitating systems
to find the $\bm{\mathcal{E}}$ and $\bm{\mathcal{B}}$ generated by one set
of source moments, given the expressions for $\bm{\mathcal{E}}$ and
$\bm{\mathcal{B}}$ for a dual set of moments.

\subsubsection{Local-Lorentz frame of a freely falling observer}

When one introduces
the local-Lorentz frame of a freely falling observer in curved spacetime,
one necessarily specializes one's foliation: (i) The local-Lorentz foliation's
space slices are flat at first order in distance from the observer's 
world line, so its extrinsic curvature $K_{ij}$ vanishes along the 
observer's world line.  (ii) Because the observer is
freely falling, her acceleration $a_k$ vanishes, which means that 
successive hypersurfaces in the foliation are parallel to each other
along the observer's world line.  

These specializations, plus the vanishing shift $\beta_i=0$ and unit lapse function $\alpha=1$
of a local-Lorentz frame, bring the constraint and evolution
equations (\ref{eq:BianchiConstraint}) and (\ref{eq:Bianchi31}) into the following Maxwell-like form:
\begin{eqnarray}
\label{LLFMaxwell}
\bm\nabla \cdot \bm{\mathcal E}&=&0\;, \quad  \bm\nabla \cdot \bm{\mathcal B}=
0\;,\\
\frac{\partial\bm{\mathcal E}}{\partial t} - \left(\bm\nabla\times \bm{\mathcal B}\right)^S &=& 0\;, \quad
\frac{\partial\bm{\mathcal B}}{\partial t} + \left(\bm\nabla\times \bm{\mathcal 
E}\right)^S = 0\;. \nonumber
\end{eqnarray}
Here the superscript $S$ means ``take the symmetric part'' and the
remaining notation is the same as in the flat-spacetime Maxwell 
equations (including changing from $\bf D$ to $\bm\nabla$ for the spatial
gradient).  

\subsubsection{Weak-gravity, nearly Minkowski spacetimes}

In this paper's applications (Secs.\ \ref{WeakGravityStationary} and
\ref{sec:GWandGeneration}), we shall specialize to spacetimes
and coordinate systems that are weakly perturbed from Minkowski, and we
shall linearize in the perturbations.  In this case, the Bianchi identities
(\ref{eq:Bianchi31}) take on precisely the same Maxwell-like form 
as in a local-Lorentz frame in strongly curved spacetime, 
Eqs.\ (\ref{LLFMaxwell}). To see that this is so, note that $\beta_k$, $K_{jk}$, $a_k$, $\mathcal E_{jk}$, and $\mathcal B_{jk}$ are all first-order perturbations and that
$\alpha$ is one plus a first-order perturbation; and linearize Eqs.\ (\ref{eq:Bianchi31}) in these first-order quantities.

When the weak-gravity spacetime is also characterized by slow motion, so its 
source regions are small compared to the wavelengths of its gravitational
waves, the evolution equations control how the near-zone 
$\mathcal E_{jk}$ and $\mathcal B_{jk}$ get transformed into gravitational-wave
fields.  
For insight into this, we specialize to harmonic gauge, in which the 
trace-reversed metric perturbation $\bar h_{\mu\nu}$ is divergence-free,
$\partial^\mu \bar{h}_{\mu \nu}=0$.  

Then \textit{in the near zone}, 
$\mathcal E_{jk}$ and $\mathcal B_{jk}$ [which are divergence-free and
curl-free by Eqs. (\ref{LLFMaxwell})] are expressible in terms of
the metric perturbation itself as
\begin{equation} \label{eq:NearZoneEB}
\mathcal{E}_{ij} = -\frac{1}{2}\partial_i \partial_j h_{00} \,, \quad \quad \mathcal{B}_{ij} =  \frac{1}{2} \epsilon_i^{\phantom{i}pq} \partial_q \partial_j h_{p0} \,.
\end{equation}
Because $h_{00}$, at leading order in $r/\lambdabar$ (ratio of radius to
reduced wavelength), contains only mass multipole moments (Eq.\ (8.13a) of
\cite{thorne80}), so also 
$\mathcal E_{jk}$ contains only mass multipole moments. And because
$h_{p0}$ at leading order in $r/\lambdabar$ contains only current
multipole moments, so also $\mathcal B_{jk}$ contains only current 
multipole moments. 

\textit{In the wave zone,} by contrast, 
Eqs.\ (\ref{LLFMaxwell}) show that
the locally plane waves are sustained by mutual 
induction between $\bm {\mathcal{E}}$ and 
$\bm{\mathcal{B}}$, just like for electromagnetic waves, which means
that these two wave-zone fields must contain the same information.
This is confirmed by
the wave-zone expressions for $\mathcal E_{jk}$ and $\mathcal B_{jk}$
in terms of the metric perturbation,
\begin{equation} \label{eq:FarZoneEB}
\mathcal{E}_{ij} =-\frac{1}{2} \partial^2_0 h_{ij} \,, \quad \quad \mathcal{B}_{ij} = -\frac{1}{2}\epsilon_i^{\phantom{1}pq} n_p \partial^2_0 h_{qj} \;.
\end{equation}
Both fields are expressed in terms of the same quantity,
$h_{ij}$. In addition, in the wave zone, $\bm {\mathcal{E}}$ and $\bm{\mathcal{B}}$ are related to each other through a $\pi/4$ rotation of their polarization tensors (see Sec.\ \ref{sec:PlaneWave} below). 
Correspondingly, we will see in Sec.\ \ref{sec:GWandGeneration} that, if a time-varying mass moment produces $+$ polarized radiation in the wave zone, then the current moment that is dual to it produces $\times$ polarized radiation of the same magnitude.

{\it In the transition zone,} the inductive coupling between 
$\bm {\mathcal{E}}$ and
$\bm{\mathcal{B}}$, 
embodied in Eqs.\ (\ref{eq:Bianchi31}), enables these equations to act like 
a blender, mixing up the multipolar information that in the near zone
is stored separately in these two fields. After an infinite amount of inductive 
blending, we arrive at future null infinity, $\mathcal{I}^+$, where the mixing has been so thorough that $\bm{\mathcal{E}}$ and $\bm{\mathcal{B}}$ contain 
precisely the same information, 
though it is distributed differently among their 
tensor components [Eqs.~\eqref{eq:FarZoneEB}].

The details of this transition-zone mixing, as embodied in 
Eqs.\ (\ref{LLFMaxwell}), are in some sense the essence of gravitational-wave
generation.  We shall explore those details visually in 
Sec.~\ref{sec:GWandGeneration}
by tracking the tendex and vortex lines (introduced in Sec~\ref{sec:NewTools})
that
extend from the near zone, through the transition zone, and into the 
far zone. 

Finally, note that the duality of $\bm{\mathcal E}$ and $\bm{\mathcal B}$ 
becomes especially convenient for slow-motion systems, where we can relate 
$\bm{\mathcal E}$ and $\bm{\mathcal B}$ to source multipole moments that appear
in the weak-field near zone. 
In particular: to obtain the $\bm{\mathcal{E}}$ and $\bm{\mathcal{B}}$ 
generated by a specific current moment $\bm{\mathcal S}_\ell$,
we can simply apply the duality transformation (\ref{eq:DualityRotation})
to
the $\bm{\mathcal{E}}$ and $\bm{\mathcal{B}}$ for its dual moment, which
is the 
mass moment $\bm{\mathcal I}_\ell$, but with one caveat:  The differing
normalizations used for mass moments and current moments \cite{thorne80}
enforce the duality relation
\begin{equation}
\label{duality}
\bm{\mathcal I}_\ell \to \frac{2 \ell}{\ell +1} \bm{\mathcal S}_\ell\;, 
\qquad \bm{\mathcal S}_\ell \to - \frac{\ell +1}{2 \ell} \bm{\mathcal I}_\ell\;,
\end{equation}
when making this duality transformation; note that both transformations, 
Eqs.~(\ref{eq:DualityRotation}) and ~(\ref{duality}), must be made at once to 
arrive at the correct expressions; see Sec.~\ref{sec:GWandGeneration}.

\section{Physical Interpretations of $\mathcal E_{ij}$ and $\mathcal B_{ij}$}
\label{sec:InterpretEandB}

It is rather well-known that in vacuum\footnote{\label{fn:vacuum}
In a non-vacuum
region of spacetime, the local stress-energy tensor also contributes to 
tidal accelerations via its algebraic relation to the Ricci tensor which
in turn contributes to the Riemann tensor. In this case,
$\mathcal E_{ij}$ describes that portion of the tidal
acceleration due to the ``free gravitational field,'' i.e., the portion 
that is sourced away from the location where the tidal acceleration is measured;
and similarly for $\mathcal B_{jk}$ and differential frame dragging.
In this paper we shall ignore this subtle point and focus on tidal forces
and differential frame dragging in vacuum.} 
the electric part of the Weyl tensor, 
$\mathcal E_{ij}$, describes tidal gravitational accelerations: the relative
acceleration of two freely falling particles with separation vector $\xi^k$  
is $\Delta a^i = -{\mathcal E^i}_j \xi^j$. For this reason $\mathcal E_{ij}$
is often called the {\it tidal field}, a name that we shall adopt.

Not so well-known is the role of the magnetic part of the Weyl tensor
$\mathcal B_{jk}$ as governing differential frame dragging, i.e.\
the differential precession of inertial
reference frames: in vacuum$^{\ref{fn:vacuum}}$ a gyroscope at the tip of the 
separation vector $\xi^k$,
as observed in the local-Lorentz frame of an observer at the tail of $\xi^k$,
precesses with angular velocity $\Delta \Omega^j = {\mathcal B^j}_k \xi^k$.
For this reason, we call $\mathcal B_{jk}$ the {\it frame-drag field}.

We deduced this frame-drag role of $\mathcal B_{jk}$ 
during our research and then searched in vain for any
reference to it in the literature, while writing our Physical Review
Letter on vortexes and tendexes~\cite{OwenEtAl:2011}.  More recently we
have learned that this role of $\mathcal B_{jk}$ was known to Frank 
Estabrook and Hugo Wahlquist~\cite{Estabrook1964} 46 years ago and 
was rediscovered 
two years ago by Christoph Schmidt~\cite{Schmid2009} 
(who states it without proof). 

For completeness, in this section we shall give a precise statement and
proof of the frame-drag role of $\mathcal B_{jk}$, and a corresponding 
precise statement of the tidal-acceleration role of $\mathcal E_{jk}$.

\subsection{Physical setup}
\label{sec:PhysicalSetup}

\begin{figure}
\includegraphics[width=0.57\columnwidth]{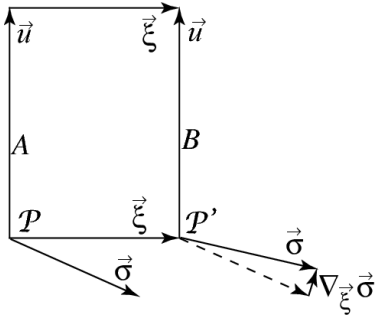}
\caption{
Spacetime geometry for computing the precession of a gyroscope at one location $\mathcal P'$,
relative to gyroscopic standards at a nearby location $\mathcal P$.
}
\label{fig:Precession}
\end{figure}

Consider an event $\mathcal P$ in spacetime and an observer labeled $A$ whose world
line passes through $\mathcal P$ and has 4-velocity $\vec u$ there; see Fig.\ \ref{fig:Precession}.
Introduce an infinitesimally short 4-vector
$\vec\xi$ at $\mathcal P$, that is orthogonal to $\vec u$ and thus is seen as spatial by observer $A$.
Denote by $\mathcal P'$ the event at the tip of $\vec \xi$.  Introduce a second observer $B$ whose
world line passes through
$\mathcal P'$ and is parallel there to the world line of observer $A$, so if we denote $B$'s 4-velocity
by the same symbol $\vec u$ as that of $A$ and imagine a vector field $\vec u$ that varies smoothly
between the two world lines, then $\nabla_{\vec \xi} \vec u = 0$ at $\mathcal P$.  Let $\vec\xi$ be transported
by observer $A$ in such a way that it continues to reach from world line $A$ to world line $B$.  Then the vectors $\vec u$ and
$\vec \xi$ satisfy
the following three relations at $\mathcal P$:
\begin{equation}
\vec\xi \cdot \vec u= 0\;, \quad
 [\vec u ,  \vec \xi] = 0, \quad \nabla_{\vec\xi} \vec u = 0\;.
\label{uxirelations}
\end{equation}
The first says that the separation vector is  purely spatial at $\mathcal P$ 
in the reference frame of observer $A$; 
the second says that $\vec \xi$ continues to reach between 
world lines $A$ and $B$, so the quadrilateral formed by $\vec u$ and $\vec \xi$ in Fig.\ \ref{fig:Precession} is closed; the third says that the two observers' world lines are parallel to each other at $\mathcal P$---i.e.,
these observers regard themselves as at rest with respect to each other.  

\subsection{Interpretation of $\mathcal E_{ij}$ as the tidal field}
\label{sec:GeodesicDeviation}

Let the two observers $A$ and $B$ fall freely, i.e.\ move on geodesics.  
Then for this physical setup, the equation of geodesic deviation states that [e.g.~\cite{MTW} Eq. (11.10)]
\begin{equation}
\nabla_{\vec u} \nabla_{\vec u} \vec\xi = - \mathsf R({\underline{\ \ }\,},\vec u,\vec \xi, \vec u)\;,
\label{GeodesicDeviation}
\end{equation}
where $\mathsf R$ is the Riemann tensor.  
In physical language, the left side is the acceleration $\Delta \vec a$ of observer $B$ 
at $\mathcal P'$, as measured in the local-Lorentz frame of 
observer $A$ at $\mathcal P$.  This relative acceleration is purely spatial as seen 
by observer $A$, and the right side of Eq.\ (\ref{GeodesicDeviation}) tells us 
that in spatial, 3-dimensional vector and tensor notation (and in vacuum so
$R_{\alpha\beta\gamma\delta} = C_{\alpha\beta\gamma\delta}$), it is given by
\begin{equation} 
\Delta a^j = -{R^j}_{\hat 0 k \hat 0} \xi^k = - {\mathcal E^j}_k \xi^k\;; \quad \hbox{i.e. }
\Delta \bm a = - \bm{\mathcal E}({\underline{\ \ }\,},\bm\xi)\;.
\label{TidalAcceleration}
\end{equation}
Since (as is well-known) this relative acceleration produces the Earth's tides when 
$\mathcal E_{jk}$ is caused by the moon and sun, $\mathcal E_{jk}$ is called the {\it tidal field},
and Eq.\ (\ref{TidalAcceleration}) is known as {\it the tidal-acceleration equation}.

\subsection{Interpretation of $\mathcal B_{ij}$ as the frame-drag field}
\label{sec:DifferentailFrameDrag}

Next let the two observers $A$ and $B$ in Fig.\ 
\ref{fig:Precession} be accelerated if they wish (with the same 4-acceleration $\vec a$ up to
differences proportional to $\vec\xi$), and give each of them
a spatial unit vector $\vec \sigma$ that is tied to an 
inertial-guidance gyroscope, so the following relations are satisfied:
\begin{equation}
\vec \sigma \cdot \vec u = 0\;, \quad \vec \sigma \cdot \vec \sigma = 1\;, \quad \nabla_{\vec u} \vec \sigma = (\vec a \cdot
\vec \sigma) \vec u\;, \quad \vec a \equiv \nabla_{\vec u} \vec u\;.
\label{Srelations}
\end{equation}
The first of these says that $\vec \sigma$ is purely spatial as seen in the observer's reference frame; 
the second says that $\vec \sigma$ has unit length; the third is the Fermi-Walker transport law for an
inertial-guidance gyroscope.  

The local-frame-dragging-induced rate of change of $\vec \sigma$ at $\mathcal P'$, as measured using
inertial-direction standards at $\mathcal P$, is $ \nabla_{\vec u} \nabla_{\vec \xi} \vec \sigma$.  We can write this as
\begin{eqnarray}
 \nabla_{\vec u} \nabla_{\vec \xi} \vec \sigma &=&  \nabla_{\vec \xi} \nabla_{\vec u} \vec \sigma 
 + [\nabla_{\vec u},\nabla_{\vec \xi}]\vec \sigma \nonumber\\
&=&  \nabla_{\vec \xi} \nabla_{\vec u} \vec \sigma + \mathsf R({\underline{\ \ }\,},\vec \sigma,\vec u, \vec \xi)\;,
 \label{FDRiemann}
 \end{eqnarray}
where $\mathsf R$ is the Riemann tensor and we have used the fact that $[\vec u,\vec\xi]=0$; cf.\
Eqs.\ (11.8) and (11.9) of MTW \cite{MTW}.

Evaluating the first term $\nabla_{\vec \xi} \nabla_{\vec u} \vec \sigma$ using the Fermi-Walker transport
law [the third of Eqs.\ (\ref{Srelations})] and the fact that the observers are momentarily at rest with
respect to each other [the third of Eqs.\ (\ref{uxirelations})], we bring Eq.\ (\ref{FDRiemann}) into
the form
\begin{equation}
 \nabla_{\vec u} \nabla_{\vec \xi} \vec \sigma = \mathsf R({\underline{\ \ }\,},\vec \sigma,\vec u, \vec \xi) + 
 \vec u \nabla_{\vec \xi} (\vec a \cdot \vec \sigma)\;.
 \label{FDRiemann1}
 \end{equation}
We are only interested in the spatial part of this rate of change, so we can ignore the second term
on the right side of the equation.
We switch to the 3-dimensional viewpoint of the observer at $\mathcal P$ (where
our calculation is being done) and we denote the spatial part of
 $\nabla_{\vec u} \nabla_{\vec \xi} \vec \sigma$
by $\dot {\bm \sigma}$:
\begin{equation}
\dot{\bm \sigma} \equiv \left[\nabla_{\vec u} \nabla_{\vec\xi} \vec \sigma\right]_{\hbox{project orthogonal to } \vec u}\;.
\label{dotSdef}
\end{equation}
Equation (\ref{FDRiemann1}) tells us that this rate of change is not only orthogonal to $\vec u$ (spatial) but also orthogonal to $\bm \sigma$; it therefore can be written as a rotation
\begin{equation}
\dot{\bm \sigma} =  \Delta \mathbf \Omega \times \bm \sigma\;
\label{Rotation}
\end{equation}
Here \textit{$\Delta  \mathbf \Omega$ is the frame-dragging angular velocity at $\mathcal P'$ as measured
using inertial standards at $\mathcal P$}.  
We can solve for this angular velocity $\Delta \mathbf \Omega$ by crossing $\bm \sigma$
into Eq.\ (\ref{Rotation}) and using $\bm \sigma \cdot \bm \sigma = 1$:
\begin{equation}
 \Delta \mathbf \Omega = \bm \sigma \times \dot{\bm \sigma}\;.
\label{Omega1}
\end{equation}
Inserting expression (\ref{FDRiemann1}) for $\dot{\bm \sigma}$ and switching to index notation, we obtain
\begin{equation}
\Delta \Omega_i = \epsilon_{ijk} \sigma^j {R^k}_{p\hat0 q} \sigma^p \xi^q\;.
\label{Omega2}
\end{equation}
Rewriting the Riemann tensor component in terms of the gravitomagnetic part of the Weyl tensor (in vacuum),
${R^k}_{p\hat0 q} = - {\epsilon^k}_{ps} {\mathcal B^s}_q$, performing some tensor manipulations, and noticing that because $\Delta\mathbf \Omega$ is crossed into $\bm \sigma$ when computing the precession any piece of $\Delta\mathbf \Omega$ along $\bm \sigma$ is irrelevant, we obtain
\begin{equation}
\Delta \Omega_i = \mathcal B_{ij}\xi^j\;, \quad \hbox{i.e. } \Delta\mathbf \Omega = 
\bm{\mathcal B}({\underline{\ \ }\,},\bm\xi)\;.
\label{DifferentialFrameDragging}
\end{equation}
Put in words: \textit{in vacuum the frame-dragging angular velocity at $\mathcal P'$, 
as measured using inertial directions at the adjacent event $\mathcal P$, is obtained by inserting the vector $\bm \xi$ (which
reaches from $\mathcal P$ to $\mathcal P'$) into one slot of the 
gravitomagnetic part of the Weyl tensor.}

Because of the role of $\mathcal B_{ij}$ in this {\it equation of differential frame dragging}, we
call $\mathcal B_{ij}$ the {\it frame-drag field}.

\section{Our New Tools: Tendex and Vortex Lines; Their Tendicities and Vorticities; Tendexes and Vortexes}
\label{sec:NewTools}

\subsection{Tendex lines and their tendicities; vortex lines and their vorticities}
\label{sec:Lines}

As symmetric, trace-free tensors, the tidal field $\bm{\mathcal E}$ and frame-drag field $\bm{\mathcal B}$ can each be characterized completely by its three principal axes (eigendirections) and its three associated eigenvalues.   

If $\bm p$ is a (smoothly changing) unit eigenvector of the tidal
field $\bm{\mathcal E}$ (or of the frame-drag field $\bm{\mathcal B}$), 
then the integral curves of $\bm p$
can be regarded as ``field lines'' associated with $\bm{\mathcal E}$ (or $\bm{\mathcal B}$).
For $\bm{\mathcal E}$ we call these integral curves {\it tidal tendex lines}, 
or simply {\it tendex lines}\footnote{The word tendex was coined by David Nichols.},
 because $\bm{\mathcal E}$ tidally stretches objects it encounters,
and the Latin word {\it tendere} means ``to stretch.''
For $\bm{\mathcal B}$ we call the integral curves {\it frame-drag vortex lines}, or simply {\it vortex lines},
because $\bm{\mathcal B}$ rotates gyroscopes, and the Latin word {\it vertere} means ``to
rotate.'' At each point $\mathcal P$ 
in space there are three orthogonal eigendirections of $\bm{\mathcal E}$ 
(and three of $\bm{\mathcal B}$), so through each point there pass three orthogonal tendex
lines and three orthogonal vortex lines. 

\begin{figure}
\includegraphics[width=0.9\columnwidth]{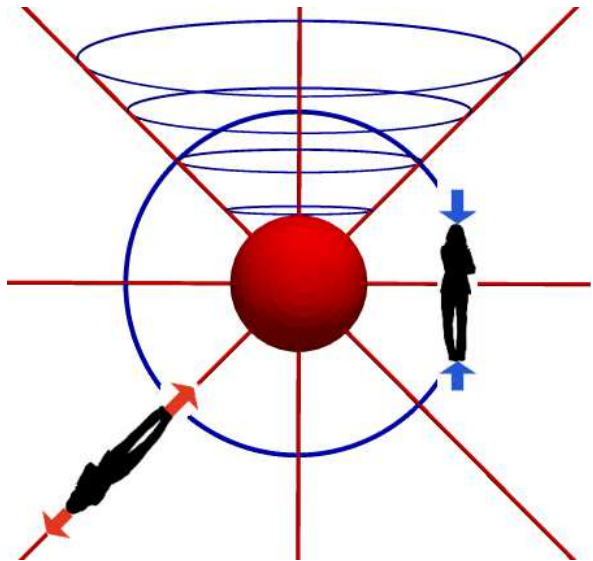}
\caption{(color online). 
Tendex lines outside a spherically symmetric, gravitating body. 
The lines are colored by the sign of their tendicity: red lines have
negative tendicity (they stretch a person oriented along them); blue lines 
have positive tendicity (they squeeze).
}
\label{fig:TendexSpherical}
\end{figure}

Outside a spherically symmetric gravitating body with mass $M$,
such as the Earth or a Schwarzschild black hole, the tidal
field, in a spherical polar orthonormal basis, has components
\begin{equation}
\mathcal E_{\hat r \hat r} = - \frac{2M}{r^3}\;, \quad \mathcal E_{\hat \theta \hat\theta}
= \mathcal E_{\hat\phi\hat\phi} = +\frac{M}{r^3}\;
\label{TidalFieldPtMass}
\end{equation}
(e.g.\ Sec.\ 1.6 and Eq.\ (31.4) of \cite{MTW}).  
The tidal-acceleration equation $\Delta a^j 
= -{\mathcal E^j}_k \xi^k$ tells us that this tidal field stretches objects
radially and squeezes them equally strongly in all tangential directions 
(see the people in 
Fig.\ \ref{fig:TendexSpherical}). Correspondingly, one eigenvector
of $\bm{\mathcal E}$ is radial, and the other two are tangential with degenerate eigenvalues.
This means that one set of tendex lines is radial (the red tendex lines in Fig.\ \ref{fig:TendexSpherical}), and any
curve lying on a sphere around the body is a tendex line.  If we break the tangential degeneracy
by picking our tangential unit eigenvectors to be the basis vectors $\bm e_{\hat \theta}$ and
$\bm e_{\hat \phi}$ of a spherical polar coordinate system, then the tangential tendex lines
are those vectors' integral curves --- the blue curves in Fig.\ \ref{fig:TendexSpherical}.

When the spherical body is weakly gravitating and is set rotating slowly, 
then it acquires a nonzero frame-drag field given by Eqs.\ (\ref{Bonespinsph})
below. The corresponding 
vortex lines are shown in Fig. \ref{fig:VortexSlowRotate}.  (See Sec.\ 
\ref{sec:OneWeakGravityBody} below for details.)

\begin{figure}
\includegraphics[width=0.9\columnwidth]{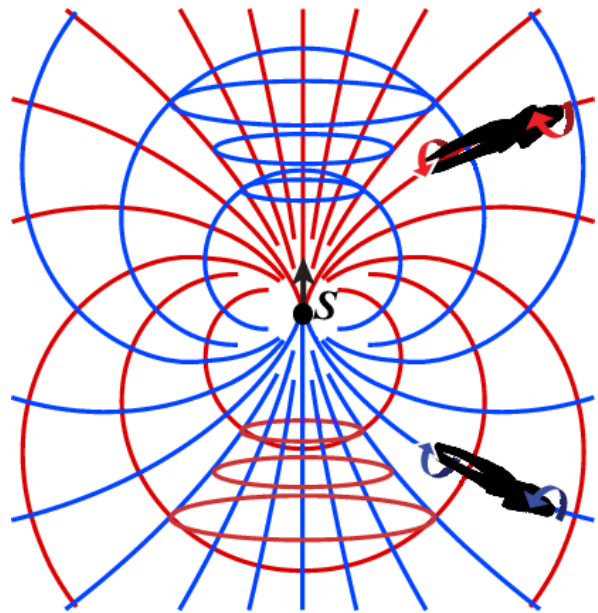}
\caption{(color online).
Vortex lines outside a slowly spinning, spherically symmetric, 
gravitating body with spin angular momentum $\bm S$.  
The lines are colored by the sign of their vorticity: red
lines have negative vorticity (they produce a counterclockwise differential
precession of gyroscopes); blue lines have positive vorticity (clockwise
differential precession).  
}
\label{fig:VortexSlowRotate}
\end{figure}

To any tendex (or vortex) line, with unit eigenvector $\bm p$, there is associated an
eigenvalue 
$\mathcal E_{pp} =  \mathcal E_{jk} p^j p^k$ which is called the line's {\it tendicity} 
(or $\mathcal B_{pp} =  \mathcal B_{jk} p^j p^k$ which is called the line's {\it vorticity}).  
The physical meaning of this tendicity (or vorticity) can be read off the 
tidal-acceleration equation (\ref{TidalAcceleration}) [or the equation of differential frame dragging
(\ref{DifferentialFrameDragging})].  Specifically, if a person's body (with length $\ell$) is oriented 
along a tidal tendex line (Fig.\ \ref{fig:TendexSpherical}), she feels a head-to-foot 
stretching acceleration 
$\Delta a = - \mathcal E_{pp} \ell$.  If the line's tendicity $\mathcal E_{pp}$ is negative
(red tendex line), her body gets stretched; if the tendicity is positive (blue tendex line),
she gets compressed.  

If her body is oriented 
along a vortex line (Fig.\ \ref{fig:VortexSlowRotate}), then a gyroscope at her feet precesses
around the vortex line with an angular speed, relative to inertial frames at her head, 
given by $\Delta\Omega = \mathcal B_{pp} \ell$.   If the line's vorticity is negative (red
vortex lines in Fig.\ \ref{fig:VortexSlowRotate}), then the gyroscope at her feet precesses 
counterclockwise relative to inertial frames at her head, and (because $\mathcal B_{pp}$ is
unchanged when one reverses the direction $\bm p$), a gyroscope at her head
precesses counterclockwise relative to inertial frames at her feet.  Correspondingly, we call 
the (red) vortex line a {\it counterclockwise vortex line}.  
If the line's vorticity is positive (blue vortex lines
in Fig.\ \ref{fig:VortexSlowRotate}), the precessions are clockwise and the vortex line is
said to be clockwise.  

For any spacetime, the tendex lines color coded by their tendicities (e.g.\
Fig.\ \ref{fig:TendexSpherical}) and the vortex lines color coded by their vorticities
(e.g.\ Fig.\ \ref{fig:VortexSlowRotate}) depict visually all details of the Weyl curvature tensor.

Since $\bm{\mathcal E}$ and $\bm{\mathcal B}$ are trace-free, at any point in space
the sum of the three tendex lines' tendicities vanishes, and the sum of the three vorticities vanishes. Because $\bm{\mathcal E}$ and $\bm{\mathcal B}$ are also symmetric,
each is characterized by five numbers at any point in space.  The direction of one tendex
line fixes two numbers and its tendicity fixes a third, leaving only two numbers
to be specified.  The direction of a second tendex line, in the plane orthogonal to the first,
fixes a fourth number and the second line's tendicity fixes the fifth and final number --- leaving
the last line's direction and tendicity fully determined.  
Similarly, this is the case for vortex lines and their vorticities.

\subsection{Vortexes and tendexes}
\label{sec:VortexesTendexes}

We give the name \textit{frame-drag vortex}, or simply \textit{vortex}, to a 
bundle of vortex lines with large vorticity. 
In Fig.\ \ref{fig:VortexSlowRotate},
the red vortex lines near the north polar axis, which
are enclosed by blue circles, 
constitute a negative-vorticity (counterclockwise)  vortex; the blue vortex lines near the south polar axis, which are enclosed by red circles,  
constitute a positive-vorticity (clockwise) vortex.  These two
vortexes emerge from the north and south poles of the spinning point particle.  

Similarly, we give the name \textit{tidal tendex}, or simply \textit{tendex}, to a
strong concentration of tendex lines.  We shall meet our first example at the end of Sec.\ 
\ref{sec:StaticMassQuadrupole} below.

\section{Weak-gravity, Stationary Systems} 
\label{WeakGravityStationary}

\subsection{One stationary, weakly gravitating, spinning body}
\label{sec:OneWeakGravityBody}

When gravity is weak and slowly changing (e.g., outside a slowly precessing, spinning,
weakly gravitating body such as the Earth), one can write the spacetime metric in the form 

\begin{subequations}
\begin{equation}
ds^2 = -\alpha^2 dt^2 + \delta_{jk} (dx^j+\beta^j dt)(dx^k+\beta^k dt)\;
\label{LTmetric}
\end{equation}
(e.g.\ Sec.\ 23.9.3 of \cite{BlandfordThorne}; 
or Chap.\ 10 of MTW \cite{MTW} with
the spatial coordinates changed slightly).  Here
\begin{equation}
\alpha^2 = \left(1-\frac{2M}{r}\right)\;, \quad \bm \beta = - \frac{2\bm S}{r^2}\times\bm n\;,
\label{lapseshift}
\end{equation}
are the squared lapse function and the shift function, $M$ is the body's mass, $\bm S$ is its 
spin angular momentum, and 
\begin{equation}
r=\sqrt{x^2+y^2+z^2}\;, \quad \bm n = \bm e_{\hat r}
\label{rn}
\end{equation}
are radius and the unit radial vector, with $\{x^1,x^2,x^3\}=\{x,y,z\}$.  In spherical polar coordinates
(associated with the Cartesian coordinates $\{x,y,z\}$ in the usual way), the metric (\ref{LTmetric})
becomes
\begin{eqnarray}
ds^2 &=& -\alpha^2 dt^2 + dr^2 + r^2 d\theta^2 + r^2 \sin^2 \theta (d\phi-\omega dt)^2\;, \nonumber\\
\omega &=& 2S/r^3\;.
\label{LTmetric_sph}
\end{eqnarray}
\end{subequations}

It is conventional to rewrite general relativity, in this weak-field, slow-motion situation, as a field
theory in flat spacetime.  In this language, the 
geodesic equation for a test particle takes the form 
\begin{equation}
\frac{d^2\bm x}{dt^2} = \bm g + \bm v\times\bm H\;,
\label{geodesicGM}
\end{equation}
which resembles the Lorentz force law in electromagnetic theory;
see, e.g., \cite{Kaplan:2009} and references therein, especially
\cite{Forward:1961}.  Here $\bm v=d\bm x/dt$ is
the particle's velocity [Cartesian components $(dx/dt,dy/dt,dz/dt)$] and 
\begin{eqnarray}
\bm g &=& -\frac12 \bm\nabla \alpha^2 = -\frac{M}{r^2} \bm n\;, \nonumber\\
\bm H &=& \bm\nabla \times \bm \beta = 2\left[\frac{\bm S - 3 (\bm S\cdot\bm n)\bm n}{r^3}\right]
\label{gH}
\end{eqnarray}
are the body's \textit{gravitoelectric field} (same as Newtonian gravitational acceleration) and its
\textit{gravitomagnetic field}.  Note that these fields have the same monopole and dipole structures as the
electric and magnetic fields of a spinning, charged particle.

In this paper we shall adopt an alternative to this ``gravito-electromagnetic'' viewpoint.  
For the gravitational influence of the mass $M$, we shall return to the Newtonian viewpoint of a gravitational acceleration $\bm g$ and its gradient, the tidal gravitational field (the electric part of the Weyl tensor) 
\begin{equation}
\bm{\mathcal E} = - \bm\nabla \bm g\;, \quad {\rm i.e.,} \; \mathcal E_{ij} 
= - g_{i,j} = \Phi_{,ij} =  \frac12 {\alpha^2}_{,ij}\;.
\label{EijNewton}
\end{equation}
Here the comma denotes partial derivative (actually, the gradient in our Cartesian coordinate system) and $\Phi$ is the Newtonian gravitational potential, which is related to the lapse
function by $\alpha^2 = 1+2\Phi$ in the Newtonian limit.  The components of this tidal field in the spherical coordinates' orthonormal basis
$\bm e_{\hat r} = \partial/\partial r$, $\bm e_{\hat \theta} = (1/r) \partial/\partial\theta$, 
$\bm e_{\hat\phi} = (1/r\sin\theta) \partial/\partial\phi$ are easily seen 
to be  
\begin{equation}
\mathcal E_{\hat r \hat r} = - \frac{2M}{r^3}\;, \quad \mathcal E_{\hat \theta \hat\theta}
= \mathcal E_{\hat\phi\hat\phi} = +\frac{M}{r^3}\;,
\end{equation}
[Eqs.\ (\ref{TidalFieldPtMass}) above],
which are symmetric and trace-free as expected.  The field lines associated with this tidal
field are easily seen to be those depicted in Fig.\ \ref{fig:TendexSpherical} above.

\begin{figure}
\includegraphics[width=0.95\columnwidth]{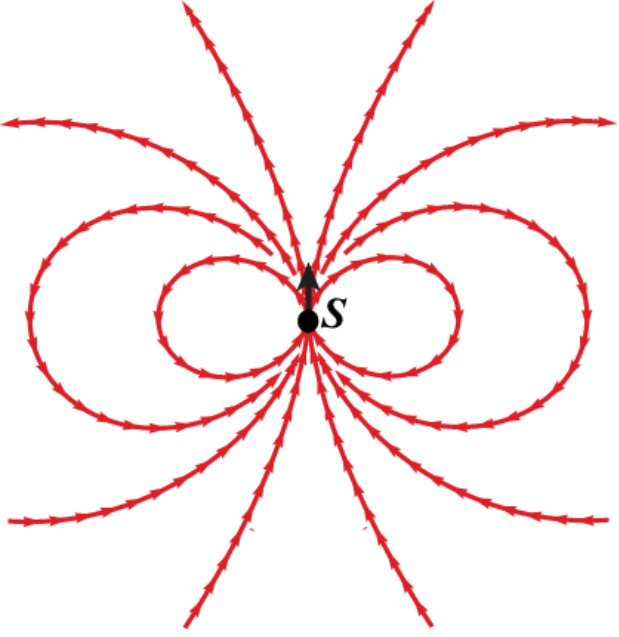}
\caption{
For a weakly gravitating, slowly
rotating body with spin angular momentum $\bm S$, 
the dipolar frame-dragging angular velocity relative to inertial frames at infinity,
$\bm \Omega_{\rm fd}$. 
The arrows are all drawn
with the same length rather than proportional to the magnitude of $\bm\Omega_{\rm fd}$.
} 
\label{fig:OmegafdOneSpin1}
\end{figure}

For the effects of the spin angular momentum, 
we shall think of the
spinning body as ``dragging space into motion'' with a velocity and angular velocity (relative to
our Cartesian coordinates) given by 
\begin{equation}
\frac{d\bm x_{\rm space}}{dt} \equiv \bm v_{\rm space} = - \bm \beta =\frac{2\bm S}{r^2}\times\bm n;, \quad
\frac{d\phi_{\rm space}}{dt} = \omega = \frac{2S}{r^3}\;
\label{vspace}
\end{equation}
[cf.\ the $\delta_{jk}(dx^j+\beta^j dt)(dx^k+\beta^k dt)$ term in the metric (\ref{LTmetric})
and the $(d\phi-\omega dt)^2$ term in the metric (\ref{LTmetric_sph})].
Just as the vorticity $\bm\nabla\times\bm v$ of a nonrelativistic fluid with velocity field $\bm v(\bm x)$
is twice the angular velocity $\bm\Omega$ of rotation of a fluid element relative to an inertial
reference frame, so 
the vorticity associated with the ``space motion,'' $\bm\nabla \times \bm v_{\rm space}$, 
turns out to be twice the
vectorial angular velocity of an inertial-guidance gyroscope relative to inertial reference frames
far from the body (``at infinity'') --- or equivalently, relative to our spatial Cartesian coordinates $\{x,y,z\}$, which
are locked to inertial frames at infinity.  In formulas: Let $\bm \sigma$ be 
a unit vector along the spin angular momentum vector 
of an inertial-guidance gyroscope.  Viewed as a vector in our Cartesian basis, it precesses 
\begin{equation}
\frac{d\bm \sigma}{dt} = \bm \Omega_{\rm fd}\times \bm\sigma\;,
\label{Sprecesss}
\end{equation}
with a {\it frame-dragging vectorial angular velocity} equal to half the vorticity of space viewed as a fluid:
\begin{eqnarray}
\bm \Omega_{\rm fd} &=& \frac12 \bm\nabla\times\bm v_{\rm space} = -\frac12 \bm\nabla\times\bm\beta
= -\frac12 \bm H
\nonumber\\
\; &=& -\left[\frac{\bm S - 3 (\bm S\cdot\bm n)\bm n}{r^3}\right];
\label{Omegafd}
\end{eqnarray}
see e.g.\  Eq.\ (25.14) of \cite{BlandfordThorne}, 
or Eq.\ (40.37) of \cite{MTW}.  This dipolar frame-dragging angular
velocity is shown in Fig.\ \ref{fig:OmegafdOneSpin1}.

For dynamical black holes and other strong-gravity, dynamical situations, it is not possible to 
measure gyroscopic precession with respect to inertial frames at infinity,
since there is no unambiguous way to compare vectors at widely separated events.\footnote{There 
is an
exception:  One can introduce additional geometric structure, e.g,\ an auxiliary flat spacetime,
that provides a way of carrying a reference frame inward from infinity to all other
locations and thereby compare vectors at different events.  
Some of us have used this approach to localize linear momentum in the
gravitational field around black holes \cite{Chen2009,Lovelace:2009}.  
However, the auxiliary
structure has great arbitrariness, and for the vortex and tendex concepts of this paper
there is no need for such auxiliary structure, so we eschew it.}

On the other hand, we {\it can}, in general, measure the precession of inertial-guidance gyroscopes
at one event, with respect to inertial frames at a neighboring event --- i.e., we can measure
{\it differential frame dragging} as embodied in the frame-drag field (magnetic part of the Weyl tensor)
$\mathcal B_{ij}$.  In our weak-gravity, slow-motion situation, this frame-drag field is equal
to the gradient of $\bm\Omega_{\rm fd}$ (Eq.\ (5.45b) of 
\cite{Thorne-Price-MacDonald:Kipversion}):
\begin{equation}
\bm{\mathcal B} = \bm\nabla \bm\Omega_{\rm fd}\;, \quad \hbox{i.e. } \mathcal B_{jk} = \Omega_{{\rm fd}\,j,k}\;.
\label{BgradOmegafd}
\end{equation}

For our weakly gravitating, spinning body, $\bm \Omega_{\rm fd}$ has the dipolar form 
(\ref{Omegafd}), so the frame-drag field is
\begin{equation}
\mathcal B_{jk} = \frac{3}{r^4}\left[ 2 S_{(j}n_{k)} +  (\bm S\cdot\bm n) 
(\delta_{jk} - 5 n_j n_k) \right]\;.
\label{Bonespin}
\end{equation}
Here the parentheses on the subscripts indicate symmetrization.  
In spherical polar coordinates, the components of this frame-drag field are
\begin{eqnarray}
\mathcal B_{\hat r\hat r} &=& -2 \mathcal B_{\hat\theta\hat\theta}=-2\mathcal B_{\hat\phi\hat\phi} = 
-\frac{6S\cos\theta}{r^4}\;, \nonumber\\
 \mathcal B_{\hat r\hat\theta} &=& \mathcal B_{\hat\theta\hat r}= -\frac{3S\sin\theta}{r^4}\;.
\label{Bonespinsph}
\end{eqnarray}

For this (and any other axially symmetric) frame-drag field, 
one of the three sets of vortex lines is along the $\phi$ direction (i.e.\ the $\bm S \times \bm x$
direction)---i.e., it is {\it axial}---and the other two are {\it poloidal}.  
By computing
the eigenvectors of the tensor (\ref{Bonespin}) and then drawing the curves to which they
are tangent, one can show that the body's vortex lines have the forms shown in
Fig.\ \ref{fig:VortexSlowRotate}.  

Notice that the poloidal, negative-vorticity vortex lines (the 
poloidal red curves in 
Fig.\ \ref{fig:VortexSlowRotate}) all emerge from the north polar region of the spinning
body, encircle the body, and return back to the north polar region.  

Why do these have negative
rather than positive vorticity?  Choose the eigendirection $\bm p$ at the body's north pole to 
point away from the body.  The body drags inertial frames in a right-handed manner (counterclockwise as seen looking down on the north 
pole), and the
frame dragging is stronger at the tail of $\bm p$ (nearer the body) than at the tip, so the 
frame-dragging angular velocity decreases from tail to tip, which means it is more left-handed (clockwise) 
at the tip than the tail; it has negative vorticity.

The poloidal, positive-vorticity vortex lines (the poloidal blue 
curves in 
Fig.\ \ref{fig:VortexSlowRotate})
all emerge from the body's south polar region, swing around 
the body, and return to the south polar region.  

The azimuthal vortex lines have negative vorticity above the hole's equatorial plane (blue azimuthal
circles) and positive vorticity below the hole's equatorial plane 
(red azimuthal circles).

\subsection{Two stationary, weakly gravitating, spinning point particles with opposite spins}
\label{TwoWeakGravityBodies}

\begin{figure}
\includegraphics[width=0.90\columnwidth]{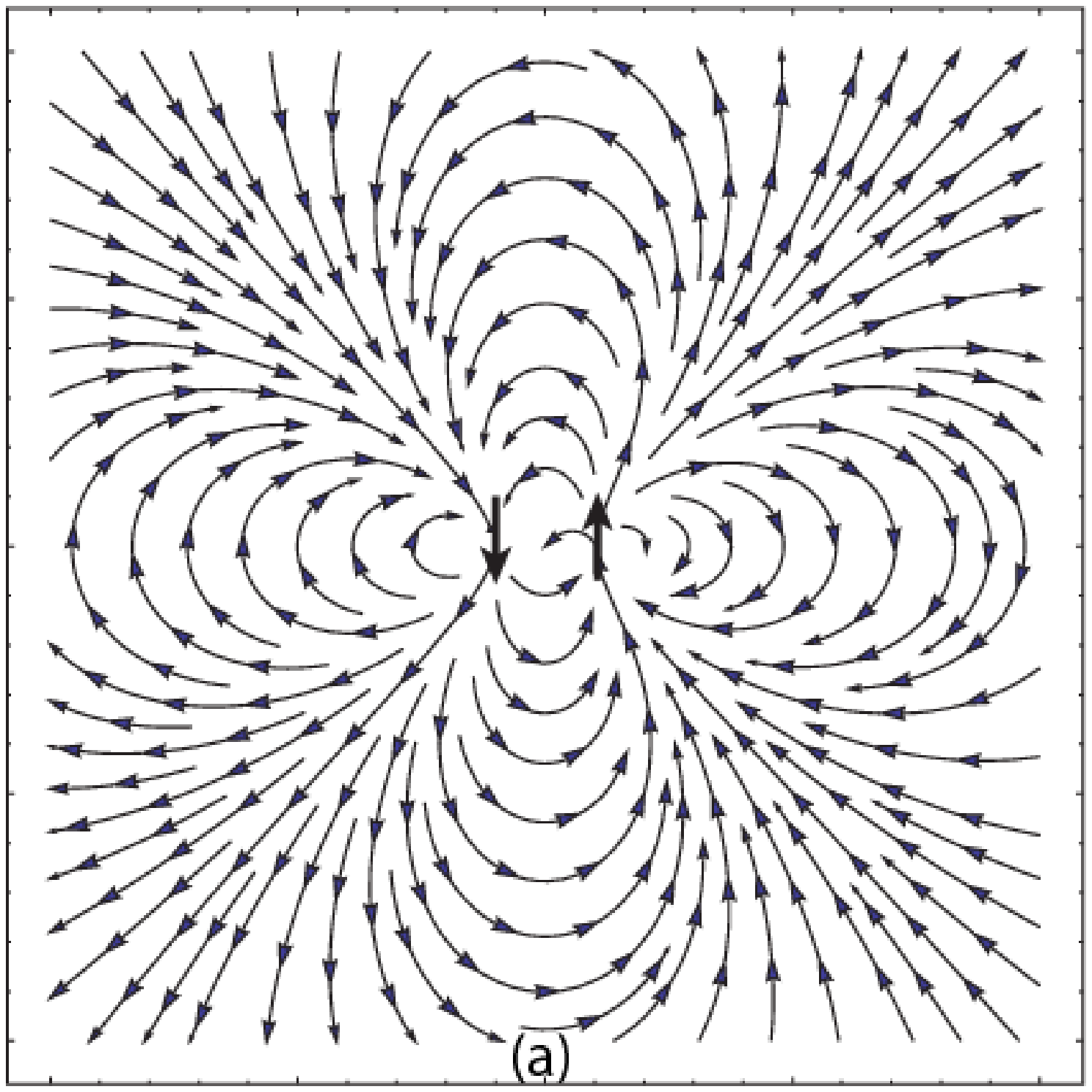}
\includegraphics[width=0.90\columnwidth]{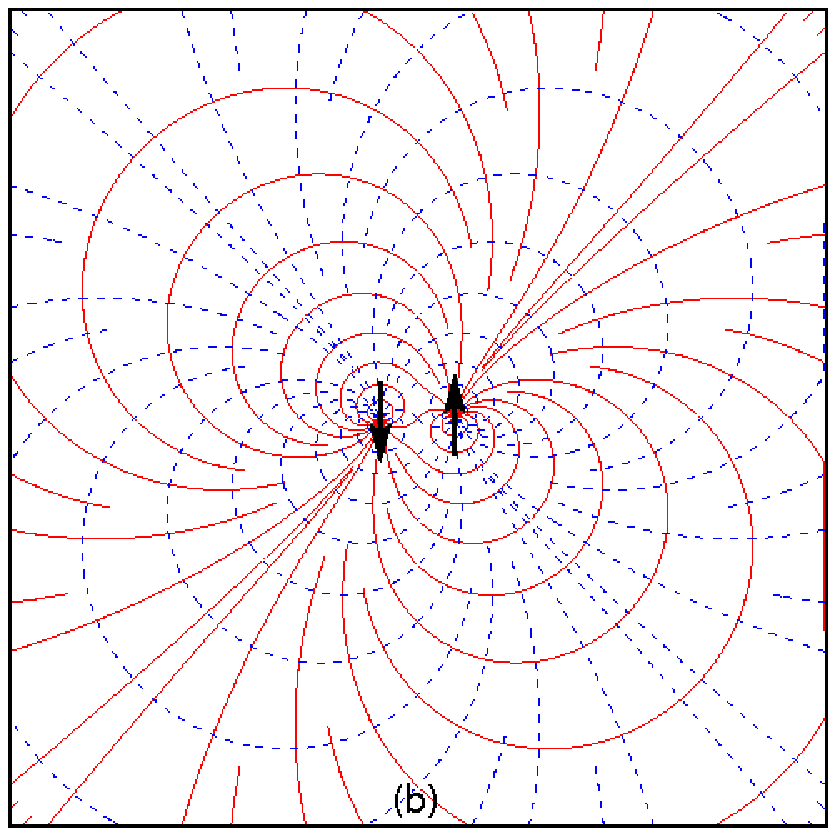}
\caption{
For two stationary point particles sitting side-by-side with their
spins in opposite directions (thick black arrows), two types of
streamlines in the plane of reflection symmetry formed by the 
particles' spins and their separation vector.
(a) The frame-dragging angular velocity $\bm\Omega_{\rm fd}$ and
its streamlines, with the arrows all drawn
at the same length rather than proportional to the magnitude of $\bm\Omega_{\rm fd}$. 
(b): The two sets of 
vortex lines of the frame-drag field $\bm {\mathcal B}$. 
The negative-vorticity vortex lines are solid and colored red, 
and the positive-vorticity ones are dashed and blue. In this figure, as in preceding figures, 
the colors are not weighted by the lines' vorticities, but only by 
the signs of the vorticities.  
 }
\label{fig:OmegafdTwoSpin}
\end{figure}

Consider, next, two weakly gravitating, spinning point particles  
with opposite spins, sitting side-by-side.
Place the particles (named $A$ and $B$) 
on the  $x$ axis, at locations $\{x_A,y_A,z_A\} = \{+a,0,0\}$, $\{x_B,y_B,z_B\} = \{-a,0,0\}$ and give them vectorial spins $\bm S_A = S \bm e_z$, $\bm S_B = - S \bm e_z$.
Then the frame-drag angular velocity relative to inertial frames at infinity  
is 
\begin{eqnarray}
\bm\Omega_{\rm fd} &=& 
-\frac{\bm S_A - 3 (\bm S_A\cdot\bm n_A)\bm n_A}{{r_A}^3}  \nonumber\\
&&-\frac{\bm S_B - 3 (\bm S_B\cdot\bm n_B)\bm n_A}{{r_B}^3}\;,
\label{OmegaTwoSpins}
\end{eqnarray}
where $r_A = | \bm x - \bm x_A|$ and $r_B = |\bm x-\bm x_B|$ are the distances to the particles and
$\bm n_A = (\bm x-\bm x_A)/r_A$ and $\bm n_B = (\bm x - \bm x_B)/r_B$ are unit vectors pointing
from the particles' locations to the field point; cf.\ Eq.\ (\ref{Omegafd}).  This vector field is plotted in Fig.\ \ref{fig:OmegafdTwoSpin}(a).
It has just the form one might expect from the one-spin field of Fig.\ \ref{fig:OmegafdOneSpin1}.

For these two spinning particles, the frame-drag field (gradient of Eq.\ (\ref{OmegaTwoSpins})] is
\begin{eqnarray}
\mathcal B_{jk} 
&=& \frac{3}{{r_A}^4}\left[ 2 S_A^{(j}n_A^{k)} +  (\bm S_A\cdot\bm n_A) (\delta^{jk} - 5 n_A^j n_A^k)\right]\; \nonumber \\
&  +&\frac{3}{{r_B}^4}\left[ 2 S_B^{(j}n_B^{k)} +  (\bm S_B\cdot\bm n_B) (\delta^{jk} - 5 n_B^j n_B^k)\right]\;\nonumber\\
\label{Btwospin}
\end{eqnarray}
[cf.\ Eq.\ (\ref{Bonespin})], where we have moved the vector and tensor indices up for simplicity
of notation. (In our Cartesian basis, there is no difference between up and down indices.)

The best two-dimensional surface on which to visualize vortex lines
of this $\bm{\mathcal B}$ is
the $x$-$z$ plane (the plane formed by the particles' spins and their separation vector).  The
system is reflection symmetric through this plane.  On this plane, one of the principal directions 
of $\bm{\mathcal B}$ is orthogonal to it (in the $y$ direction); the other two lie in the plane and
are tangent to the in-plane vortex lines.  By computing the eigendirections
of $\bm{\mathcal B}$ [i.e., of the tensor (\ref{Btwospin})] and mapping out their tangent vortex lines,
and checking the sign of $\mathcal B_{pp}$ along their tangent directions $\bm p$, we obtain
Fig.\ \ref{fig:OmegafdTwoSpin}.

Note that, as for a single spinning particle (Fig.\ \ref{fig:VortexSlowRotate})
, so also here for two spins, the negative-vorticity vortex lines (solid red 
curves) emerge from the tips of the spins and the positive-vorticity vortex 
lines (dashed blue curves) emerge from their tails.  For a single spin,
the negative-vorticity vortex lines emerge from the tip, travel around the 
body, and return to the same tip.  
Here, the lines close to each spinning body leave and enter the same body's
tip, but the majority emerge from one body's tip, travel around that body and 
enter the other body's tip. 
Similarly the positive-vorticity vortex lines (dashed and blue) emerge from 
one body's tail, travel around that body, and enter the other body's tail
(aside from the lines near each body that exit and return to the
same body's tail).

The collection of solid red vortex lines near each arrow tip in Fig.\ \ref{fig:OmegafdTwoSpin}(b) constitutes a  negative-vorticity
frame-drag vortex, and the collection of dashed blue vortex lines near each arrow tail is a positive-vorticity vortex.  

\begin{figure} [t!]
\includegraphics[width=0.90\columnwidth]{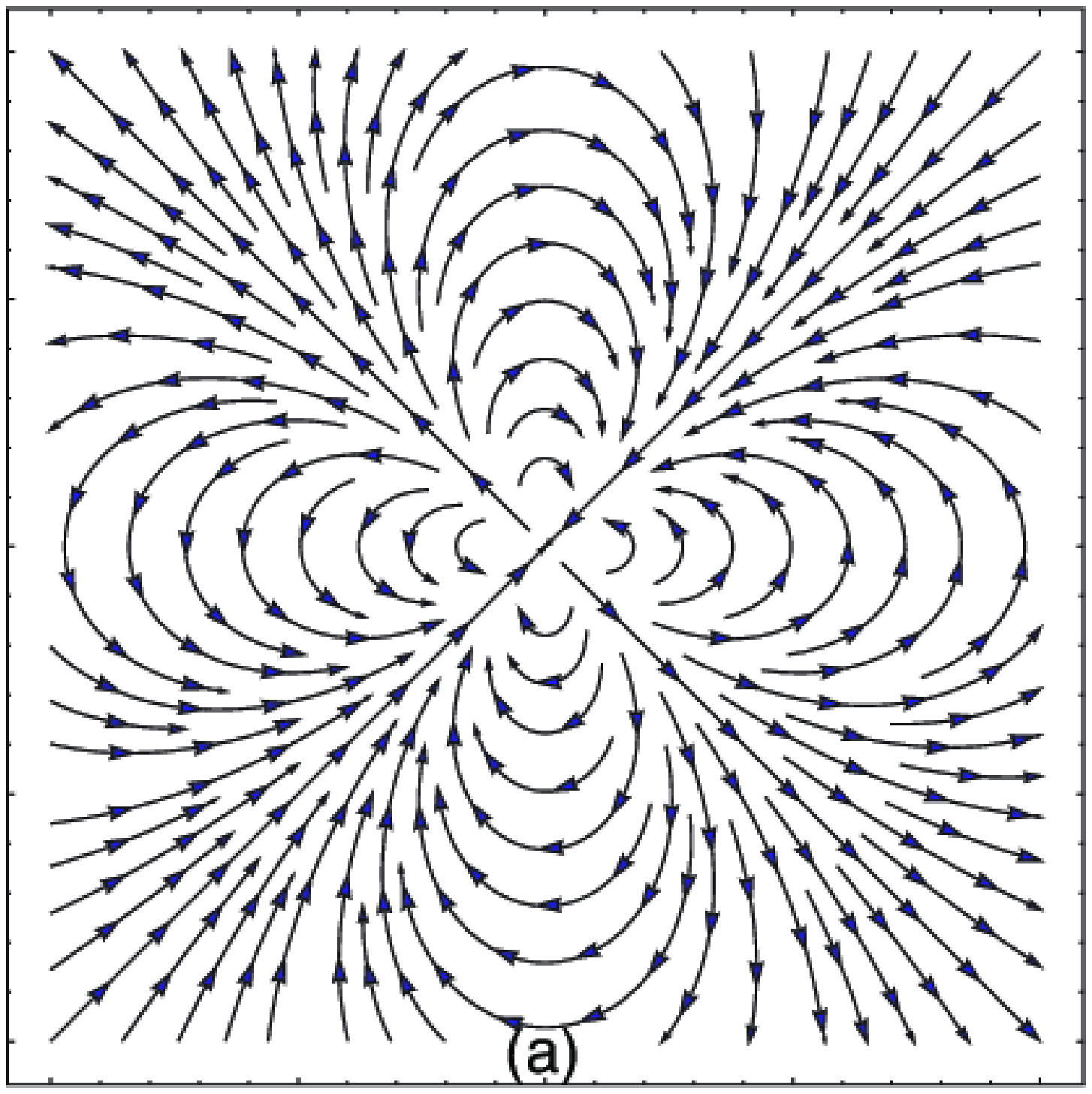}
\includegraphics[width=0.90\columnwidth]{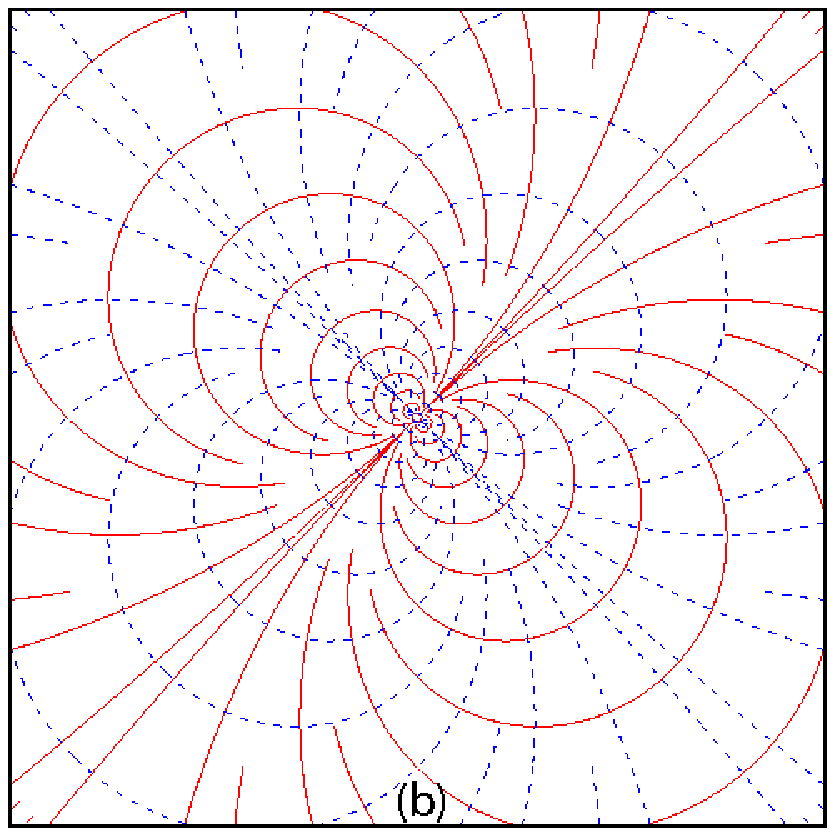}
\caption{
Current-quadrupolar streamlines associated with the two 
stationary spinning particles of Fig.\ \ref{fig:OmegafdTwoSpin},
for which the current-quadrupole moment has nonzero components
$\mathcal S_{xz} = \mathcal S_{zx} = Sa$. 
(a) The frame-dragging angular velocity $\bm\Omega_{\rm fd}$ and
its streamlines, and (b) the two
sets of 
vortex lines, in the $x$-$z$ plane.
Figure (b) also describes the tendex lines
for a static mass-quadrupolar particle whose only 
nonzero quadrupole-moment components are 
$\mathcal I_{xz} = \mathcal I_{zx}$.  
 }
\label{fig:OmegafdSpq}
\end{figure}

\subsection{The two spinning particles viewed from afar: Stationary, quadrupolar frame-drag field}
\label{sec:StationaryCurrentQuadrupole}

When viewed from afar, the two spinning bodies produce a current-quadrupole gravitational
field with quadrupole moment  (e.g.\ Eq.\ (5.28b) of \cite{thorne80})
\begin{eqnarray}
\mathcal S_{pq} &=& \left( \int  j_p x_q d^3x \right)^{\rm STF} = \left(S_p a_q + (-S_p)(-a_q)\right)^{\rm STF}    \nonumber \\
&=& S_p a_q + S_q a_p - \frac23 (\bm S \cdot \bm a) \delta_{pq}\:.
\label{Spq}
\end{eqnarray}
Here $j_p = S_p \delta(\bm x - \bm a)  - S_p \delta(\bm x + \bm a)$ is the angular momentum density.
Since the only nonzero components of $\bm S$ and $\bm a$ are $S_z = S$ and $a_x=a$, the only nonzero
components of the current-quadrupole moment are
\begin{equation}
\mathcal S_{xz} = \mathcal S_{zx} = S a\;.
\label{Sxz}
\end{equation}
The frame-drag-induced velocity of space (negative of the shift function) for this current quadrupole,
and the frame-drag angular velocity and frame-drag tensor field are
\begin{eqnarray}
\bm v_{\rm space} &=& -\bm \beta = \frac{4 \bm n \times \bm{\mathcal S}\cdot \bm n}{r^3}\;, 
\nonumber \\
\bm\Omega_{\rm fd} &=& \frac12 \bm\nabla \times \bm v_{\rm space}\;, \quad
\bm{\mathcal B} = \bm\nabla \bm\Omega_{\rm fd}\;.
\label{vOmegaBSpq}
\end{eqnarray}
[e.g.\ Eq.\ (10.6b) of \cite{thorne80}; also Eqs.\ (\ref{Omegafd}) and (\ref{BgradOmegafd}) above].
Inserting Eq.\ (\ref{Sxz}) for the quadrupole moment into Eqs.\ (\ref{vOmegaBSpq}), and  
plotting $\bm \Omega_{\rm fd}$ and the vortex lines of $\bm{\mathcal B}$ in the $x$-$z$
plane, we obtain the graphs shown in Fig.\ \ref{fig:OmegafdSpq}.

Notice that the current-quadrupolar frame-drag angular velocity in Fig.\ \ref{fig:OmegafdSpq}(a)
is, indeed, the same as that for two oppositely directed spins [Fig.\ \ref{fig:OmegafdTwoSpin}(a)] in
the limit that the spins' separation goes to zero---i.e., as seen from 
afar---and the current-quadrupolar
vortex lines of the frame-drag tensor field [Fig.\ \ref{fig:OmegafdSpq}(b)] is the vanishing-separation
limit of that for the two oppositely directed spins (Fig.\ \ref{fig:OmegafdTwoSpin}b).

Here, as for finitely separated spinning particles, there are two red frame-drag vortexes, one emerging 
from the origin in the upper right direction, the other in the lower-left direction; and similarly,
there are two blue frame-drag vortexes, one emerging in the upper left direction and the other
in the lower right direction.  

\subsection{Static, quadrupolar tidal field \\ 
and its tendex lines and tendexes}
\label{sec:StaticMassQuadrupole}

For an idealized static particle with time-independent 
mass-quadrupole moment $\mathcal I_{pq}$ and all other
moments (including the mass) vanishing, the squared lapse function is 
$\alpha^2 = 1+2\Phi = 1 - (\mathcal I_{pq}/r)_{,pq}$ \cite{thorne80}, where  $\Phi$ is the Newtonian gravitational
potential.  Therefore, the particle's tidal field $\mathcal E_{jk} = \Phi_{,jk}$ [Eq.\ (\ref{Omegafd})] is
\begin{equation}
\mathcal E_{jk} = -\frac12 \left( \frac{\mathcal I_{pq}}{r}\right)_{,pqjk}\;.
\label{MassQuadrupoleE}
\end{equation}
For comparison, for a particle with time-independent current-quadrupole moment
$\mathcal S_{pq}$, the shift function is
$ \beta_j = (-4/3) \epsilon_{jpq} (\mathcal S_{pk}/r)_{,kq}$, which implies that the frame-drag
field is [Eqs.\ (\ref{vOmegaBSpq})]
\begin{equation}
\mathcal B_{jk} = -\frac23 \left(\frac{\mathcal S_{pq}}{r}\right)_{,pqjk}\;.
\label{CurrentQuadrupoleB}
\end{equation}

Notice that, once the differing normalization conventions 
(\ref{duality}) are accounted for, Eqs.\ (\ref{MassQuadrupoleE}) and 
(\ref{CurrentQuadrupoleB}) are the same, as required by the duality relations 
\eqref{eq:DualityRotation} and (\ref{duality}).
This means that, for a static current quadrupole whose only
nonzero components are $\mathcal I_{xz} = \mathcal I_{zx}$, the tendex lines will have
precisely the same forms as the vortex lines of the static current quadrupole (\ref{Sxz}); i.e., they will have the forms
shown in Fig.\  \ref{fig:OmegafdSpq}b. 
In this case there are two negative-tendicity
(solid red) tidal tendexes, one emerging from the origin in the upper right direction, and the
other in the lower-left direction; and there are two positive-tendicity (dashed blue) tidal tendexes, one emerging in the upper left direction and the other in the lower right direction.  

\section{Gravitational Waves and their Generation}
\label{sec:GWandGeneration}

We turn now to dynamical situations, which we describe using linearized gravity. We first discuss the tendex and vortex structure of plane gravitational waves. We then examine wave generation by time-varying multipolar fields, and the accompanying tendex and vortex structures of these systems.

\subsection{Plane gravitational wave}
\label{sec:PlaneWave}

In this section, we will describe the features of $\bm{\mathcal{E}}$ and $\bm{\mathcal{B}}$ for plane gravitational waves, and connect our observations to the linearized-gravity and Newman-Penrose (NP) formalisms. In Appendix \ref{sec:NPFormalism} we review the Newman-Penrose  formalism and its connection to the spatial tensors $\bm{\mathcal E}$ and $\bm{\mathcal B}$.

Consider gravitational-wave propagation in an asymptotically flat spacetime, 
in transverse-traceless (TT) gauge. Near future null infinity, $\mathcal{I}^+$, we can linearize around a Minkowski background and obtain 
\begin{equation} 
\mathcal{E}_{ij} = -\frac {1}{2} \partial^2_0 h_{ij} \,, \quad \quad \mathcal{B}_{ij} = -\frac{1}{2} \epsilon_i^{\phantom{i} pq} n_p \partial^2_0 h_{qj} \,.
\end{equation}
It is convenient to expand these expressions in terms of 
the two gravitational-wave polarization tensors, $e^+_{ij}$ and 
$e^\times_{ij}$,
\begin{equation} \label{eq:EBLag}
\begin{split}
\mathcal{E}_{ij} &= - \frac{1}{2}(\ddot{h}_{+} e^{+}_{ij} + \ddot{h}_{\times} e^{\times}_{ij}) \, ,\\
\mathcal{B}_{ij} &= - \frac{1}{2}( \ddot{h}_{+} e^{\times}_{ij} - \ddot{h}_{\times} e^{+}_{ij} ) \, ,
\end{split}
\end{equation}
where $e^+_{ij}$ and $e^\times_{ij}$ are symmetric, trace-free, and orthogonal
to the waves' propagation direction.
Letting the unit-norm vector ${\bf e}_{\hat 1}$ denote the direction of 
propagation of the gravitational wave, then one can expand the polarization 
tensors in terms of the remaining two vectors of an orthonormal triad, 
${\bf e}_{\hat 2}$ and ${\bf e}_{\hat 3}$, as 
\begin{eqnarray}
{\bm e}_{+} &=& {\bm e}_{\hat 2} \otimes {\bm e}_{\hat 2} -{\bm e}_{\hat 3} \otimes {\bm e}_{\hat 3} \label{eq:PlusBase} \,, \\ 
{\bm e}_{\times} &=&  {\bm e}_{\hat 2} \otimes {\bm e}_{\hat 3} + {\bm e}_{\hat 3}  \otimes {\bm e}_{\hat 2}\,. \label{eq:CrossBaseA}
\end{eqnarray}

Consider first a $+$ polarized wave. 
We have that
\begin{equation} \label{eq:EigenBrek}
\bm{\mathcal E} =-\frac{1}{2} \ddot{h}_{+} {\bm e}^{+} = \frac{1}{2}[ (- \ddot{h}_{+}) {\bm e}_{\hat 2} \otimes {\bm e}_{\hat 2} +\ddot{h}_{+} {\bm e}_{\hat 3} \otimes {\bm e}_{\hat 3}] \,,
\end{equation}
so we see that $\mp \ddot{h}_{+}/2$ are the two eigenvalues of 
$\bm{\mathcal E}$ (the two tendicities), 
and ${\bm e}_{\hat 2}$ and ${\bm e}_{\hat 3}$ are the two 
corresponding eigenvectors. 
Now, define a second basis locally rotated at each point by 
$\pi/4 = 45^{\rm o}$,
\begin{equation} \label{eq:BasisRot}
\bma \tilde{\bm e}_{\hat 2} \\ \tilde{\bm e}_{\hat 3} \ema = \bma \cos \frac{\pi}{4} & \sin\frac{\pi}{4} \\ - \sin\frac{\pi}{4} & \cos\frac{\pi}{4} \ema \bma {\bm e}_{\hat 2}\\ {\bm e}_{\hat 3} \ema \, .
\end{equation}
Then, a simple calculation shows that
\begin{equation} \label{eq:CrossBaseB}
{\bm e}_{\times} = \tilde{\bm e}_{\hat 2} \otimes \tilde{\bm e}_{\hat 2} - \tilde{\bm e}_{\hat 3} \otimes \tilde{\bm e}_{\hat 3} \, ,
\end{equation}
and one can immediately see that $\bm{\mathcal B}$ is diagonal 
in this new basis
\begin{equation}
 \label{eq:EigenErek}
\bm{\mathcal{B}} = - \frac{1}{2} \ddot{h}_{+}{\bm e}^{\times} = -\frac{1}{2}[ \ddot{h}_{+} \tilde{\bm e}_{\hat 2} \otimes \tilde{\bm e}_{\hat 2} -\ddot{h}_{+} \tilde{\bm e}_{\hat 3} \otimes \tilde{\bm e}_{\hat 2}] \, .
\end{equation}
The eigenvalues of $\bm{\mathcal B}$ (the vorticities), like those of 
$\bm{\mathcal E}$ (the tendicities),
are $\mp \ddot{h}_{+}/2$,
but $\bm{\mathcal B}$'s eigenvectors, 
$\tilde{\bm e}_{\hat 2}$ and $\tilde{\bm e}_{\hat 3}$,
are locally rotated by $\pi/4$ compared to those of $\bm{\mathcal E}$.
Correspondingly, the vortex lines of $h_+$ must be locally rotated by $\pi/4$
with respect to the tendex lines.

\begin{figure}
\includegraphics[width=0.95\columnwidth]{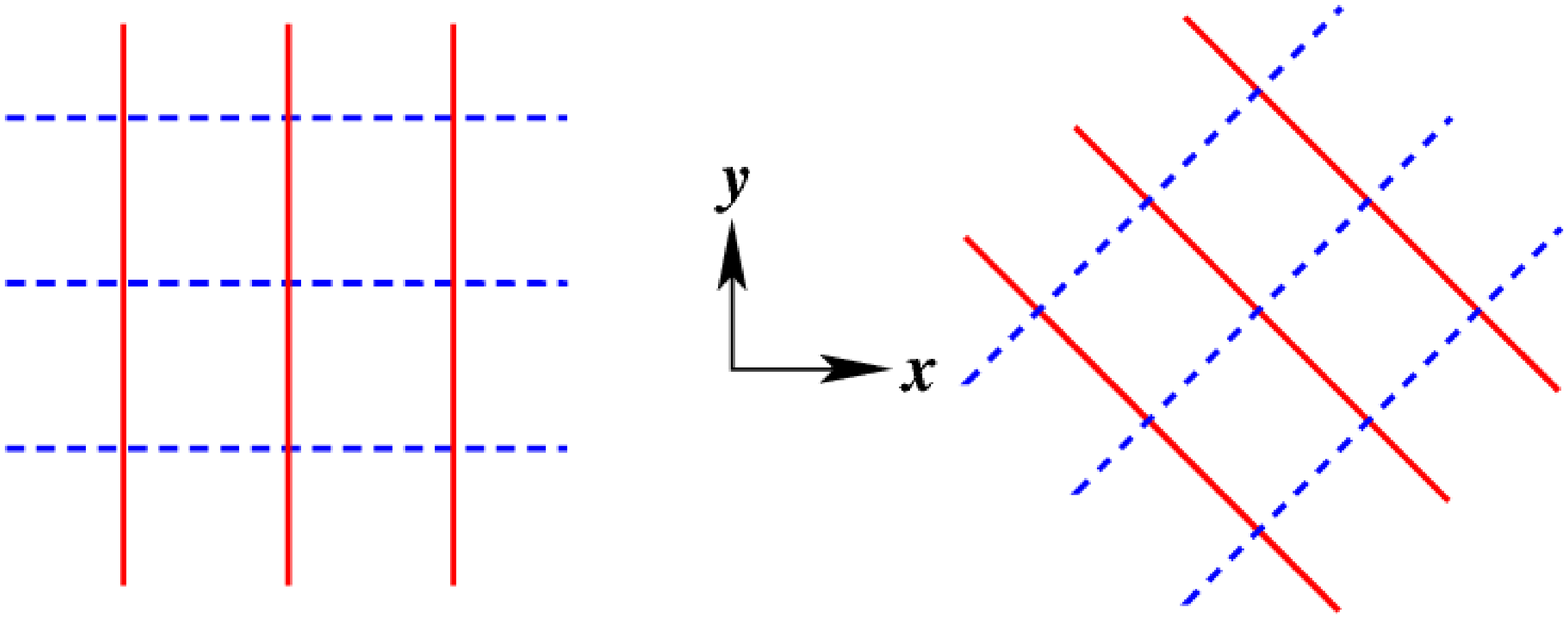}
\caption{The tendex lines (left) and vortex lines (right) 
of a plane gravitational
wave propagating in the $z$ direction (out of the picture).
The tendex lines are lines of constant $x$ and $y$, and the vortex lines
are rotated by $\pi/4$ (lines of constant $x\pm y$).
The blue (dashed) curves correspond to positive tendicity and vorticity
(squeezing and clockwise differential precessing, respectively) and
the red (solid) curves denote negative tendicity and vorticity
(stretching and counterclockwise precessing).
The tendicity (vorticity) is constant along a tendex line (vortex line), 
and the tendicity (vorticity) of a red line is equal in magnitude but
opposite in sign to that of a blue (dashed) line.} 
\label{fig:PlaneWave}
\end{figure}

The local rotation of the tendex and vortex lines is most transparent
for a plane gravitational wave.
In Fig.\ \ref{fig:PlaneWave}, we show the tendex and vortex lines of a plane 
gravitational wave propagating out of the page 
(i.e.\ ${\bf e}_{\hat 1} = {\bf e}_{\hat z}$ is the propagation direction).
Because the eigenvectors of $\bm{\mathcal E}$ are 
${\bf e}_{\hat 2}={\bf e}_{\hat x}$ and ${\bf e}_{\hat 3}={\bf e}_{\hat y}$,
the tendex lines are the lines of constant $x$ and $y$, illustrated by
red (solid) lines and blue (dashed) lines, respectively, on the left of
Fig.\ \ref{fig:PlaneWave}.
Similarly, the vortex lines are lines of constant $x\pm y$, again
drawn as blue (dashed) lines and red (solid) lines, respectively.
The tendicity (vorticity) has constant magnitude along the lines, but 
the two sets of tendex (vortex) lines have opposite sign; consequently,
the tidal (frame-drag) field produces a stretching (counterclockwise 
differential precession) along the solid red direction and a squeezing 
(clockwise differential precession) of the same magnitude along the dashed 
blue direction.

More generally, gravitational waves will contain both $+$ and
$\times$ polarizations, and to study their vortex and tendex lines, it
will be useful to express
the electric and magnetic tensors in the spatial orthonormal basis 
$({\bm e}_{\hat 1},{\bm e}_{\hat 2},{\bm e}_{\hat 3})$.
They can be written conveniently as matrices:
\begin{subequations}
\label{eq:EBh}
\begin{eqnarray}
\mathcal{E}_{\hat a \hat b} &=& \frac 12 \left(\begin{array}{c|cc}
 0 & 0 & 0  \\ \hline
0&  - \ddot{h}_{+} & - \ddot{h}_{\times}  \\ 
0& - \ddot{h}_{\times} & \ddot{h}_{+}  
 \end{array} \right) \,,
\\
\mathcal{B}_{\hat a \hat b} &= & \frac 12 \left(\begin{array}{c|cc}
 0 & 0 & 0  \\  \hline
 0& \ddot{h}_{\times} & -\ddot{h}_{+}  \\ 
 0 & -\ddot{h}_{+} & -\ddot{h}_{\times} 
 \end{array} \right) \,.
\end{eqnarray}
\end{subequations}
It is useful to
introduce an associated Newman-Penrose null tetrad consisting of two real 
null vectors, $\vec l$ (along the waves' propagation direction) and $\vec n$,
and a conjugate pair of complex 
null vectors $\vec m$ and $\vec m^*$ given by 
\begin{eqnarray}
\vec l = \frac{1}{\sqrt{2}}(\vec e_{\hat 0} + \vec e_{\hat 1}) \,, & &
\vec n = \frac{1}{\sqrt{2}}(\vec e_{\hat 0} - \vec e_{\hat 1}) \,, \nonumber \\
\vec m = \frac{1}{\sqrt{2}}(\vec e_{\hat 2} + i \vec e_{\hat 3})\,,  & &
\vec m^* = \frac{1}{\sqrt{2}}(\vec e_{\hat 2} - i \vec e_{\hat 3}) \,
\label{NullTetrad1}
\end{eqnarray}
[Eqs.\ (\ref{NullTetrad}) of Appendix A]. 
For plane waves on a Minkowski background, the NP curvature scalar that characterizes the radiation is
\begin{equation}
\label{eq:PsiVh}
\Psi_4 = C_{\mu \nu \rho \sigma} n^\mu m^{* \nu} n^\rho m^{* \sigma} 
= -\ddot{h}_{+} + i \ddot{h}_{\times} \,,
\end{equation} 
so we can compactly rewrite Eqs.\  (\ref{eq:EBh}) as 
\begin{equation} \label{eq:PropaEM}
\mathcal{E}_{\hat a \hat b} + i \mathcal{B}_{\hat a \hat b} = \frac{1}{2}\left(\begin{array}{c|cc}
 0 & 0 & 0  \\  \hline
0& \Psi_4 & i\Psi_4  \\ 
0&  i\Psi_4 & -\Psi_4
 \end{array} \right) \,.
\end{equation}
This expression holds for any plane gravitational wave propagating in the 
$\vec e_{\hat 1}$ direction.  

For any outgoing gravitational wave in an asymptotically flat space, as one 
approaches asymptotic null infinity the general expression 
\eqref{eq:Q} for $\mathcal{E}_{\hat a \hat b} + i \mathcal{B}_{\hat a \hat b}$ 
reduces to expression (\ref{eq:PropaEM}), because all the 
curvature scalars except $\Psi_4$ vanish due to the peeling property of 
the Weyl scalars near null infinity. Further discussion of the tidal and frame-drag fields of radiation 
near null infinity and their tendex and vortex lines is given in \cite{Zimmerman2011}.

It is helpful to draw some simple analogies between gravitational and 
electromagnetic plane waves. For a generic mixture of $+$ and $\times$ 
polarizations, 
the magnitudes of the nonvanishing eigenvalues of both $\bm{\mathcal E}$ 
and $\bm{\mathcal B}$ are simply 
\begin{equation}
\frac{1}{2}\sqrt{\ddot{h}^2_{+}+\ddot{h}^2_{\times}} =\frac{1}{2} |\Psi_4| \,.
\end{equation}
This mirrors plane waves in electromagnetism, 
where $|\vec{E}|=|\vec{B}|$ is equal to the sum in quadrature of the 
magnitudes of the two polarizations. The absent longitudinal components 
in an electromagnetic plane wave correspond to the vanishing of the 
eigenvalues for the eigenvectors of $\bm{\mathcal E}$ and $\bm{\mathcal B}$ along the propagation direction. 
The orthogonality of the vectorial electromagnetic field strengths 
$\vec{E}\bot \vec{B}$ becomes the $\pi/4$ rotation between the meshes 
(Fig.\ \ref{fig:PlaneWave}) formed by the two transverse eigenvectors of the tensorial quantities $\bm{\mathcal E}$ and $\bm{\mathcal B}$.

\subsection{Gravitational waves from a head-on collision \\
of two black holes} 
\label{sec:GWsHeadOnCollision}

As an example of the usefulness of this approach, we calculate the 
tendex and vortex lines at large radii for gravitational waves emitted by the 
head-on collision of two equal-mass nonspinning black holes. 
If the holes move along the $z$ axis and we use as our
spatial triad the unit vectors of spherical polar coordinates,
$({\bf e}_{\hat 1},{\bf e}_{\hat 2},{\bf e}_{\hat 3}) = 
({\bf e}_{\hat r},{\bf e}_{\hat \theta},{\bf e}_{\hat \phi}) =
(\partial_r,r^{-1}\partial_\theta,(r\sin\theta)^{-1}\partial_\phi)$, 
and choose our null tetrad in the usual way (\ref{NullTetrad1}), then
we can apply the results described by Fiske et al.\ \cite{Fiske2005},
namely, that $\Re[\Psi_4]$ is axisymmetric [and, when decomposed into 
spin-weighted spherical harmonics, is dominated by the $l=2$, $m=0$ 
harmonic, ${}_{-2}Y_{2,0}(\theta,\phi)$] and that $\Im[\Psi_4]=0$.
Then the electric and magnetic parts of the
Weyl tensor are given by
\begin{equation} \label{eq:EBMatrix}
\begin{split}
\mathcal{E}_{\hat a \hat b} &= \frac{1}{2}\left(\begin{array}{c|cc}
 0 & 0 & 0  \\  \hline
 0 & \Re(\Psi_4) & 0  \\ 
 0 & 0& -\Re(\Psi_4) 
 \end{array} \right)  \,,
\\ 
\mathcal{B} _{\hat a \hat b}&= \frac{1}{2}\left(\begin{array}{c|cc}
 0 & 0 & 0  \\  \hline
 0& 0 &  \Re(\Psi_4)  \\ 
 0&  \Re(\Psi_4) & 0 
 \end{array} \right) \,,
\end{split}
\end{equation}
and the eigenvalues of both $\bm{\mathcal{E}}$ and $\bm{\mathcal{B}}$ are 
$\pm \Re(\Psi_4)/2$.
The eigenvectors of $\bm{\mathcal E}$ are the unit vectors 
${\bf e}_{\hat \theta}$ and ${\bf e}_{\hat \phi}$, and those of 
$\bm{\mathcal B}$ are ${\bf e}_{\hat \theta} \pm {\bf e}_{\hat \phi}$.
Thus, 
the radiation is purely $+$ polarized in this basis.
The tendex lines are the lines of constant $\theta$ and $\phi$ on
a sphere, and the vortex lines are rotated relative to the tendex lines
by $\pi/4 = 45^{\rm o}$.

\begin{figure}
\includegraphics[width=0.95\columnwidth]{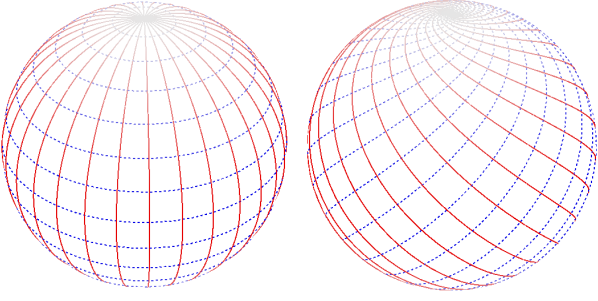}
\caption{Tendex lines (left) and vortex lines (right) for the
gravitational waves that would arise from the merger of equal-mass black
holes falling together along the z axis.
The positive tendicity and vorticity lines are shown in blue (dashed) and
the negative lines are depicted in red (solid).
Each line's intensity is proportional to its tendicity (or vorticity), which
varies over the sphere as
the dominant spin-weighted spherical harmonic,
${}_{-2}Y_{2,0}(\theta,\phi)\propto \sin^2\theta$.
Dark red and blue near the equator correspond to large-magnitude tendicity and 
vorticity, and light nearly white colors at the poles 
indicate that the tendicity and vorticity are small there. 
}
\label{fig:hPlusSphere}
\end{figure}

We show these lines in Fig.\ \ref{fig:hPlusSphere}: the tendex
lines on the left, and the vortex lines on the right. 
As in Fig.\ \ref{fig:PlaneWave}, the red (solid) lines correspond to
negative tendicity and vorticity, and the blue (dashed) lines denote
positive values.
The intensity of each line is proportional to the
magnitude of its tendicity (or vorticity), which varies 
over the sphere as 
${}_{-2}Y_{2,0}(\theta,\phi)\propto \sin^2\theta$
(the dominant spherical harmonic).
Correspondingly, the dark blue and red regions near the
equator represent strong tendicity and vorticity, whereas the light 
off-white colors near the poles indicate that the tendicity and vorticity are 
small there.

We remark in passing that the duality of $\bm{\mathcal E}$ and 
$\bm{\mathcal B}$ implies that, if there were a source of gravitational waves
which had a $\Psi_4$ that is purely imaginary and equal to the  
$i \Re[\Psi_4]$ for our colliding black holes, then those waves' vortex 
lines would be
the same as the tendex lines of Fig.\ \ref{fig:hPlusSphere}, and the tendex
lines would be the same as the vortex lines of the same figure (but with the
sign of the lines' vorticity flipped).
One can see this because (i) Eq.\ (\ref{eq:PsiVh}) shows we would have a 
pure $\times$ polarized wave, and (ii) 
when we apply the rotation of basis 
(\ref{eq:BasisRot}) to (\ref{eq:PropaEM}) under the condition of 
$\Re(\Psi_4)= 0$ we get once again the matrices (\ref{eq:EBMatrix}), but with 
$(\tilde{\bm e}_{\hat 2},\tilde{\bm e}_{\hat 3})$ as basis vectors and with 
all instances of $\Re(\Psi_4)$ replaced by $\Im(\Psi_4)$.
This duality does not address, however, how to construct a source with
a purely imaginary $\Psi_4$.

\subsection{Wave generation by a time-varying current quadrupole}
\label{sec:CurrentQuadrupoleWaveGeneration}

A dynamical current-quadrupole moment $\mathcal S_{pq}(t)$ generates a metric perturbation described by the $\mathcal S_{pq}(t-r)/r$ terms in
Eqs.\ (8.13) of \cite{thorne80}.  It is straightforward to show that the corresponding frame-drag
field is 
\begin{eqnarray}
\mathcal B_{ij} &=& \frac 23\left[-\left(\frac{\mathcal S_{pq}}{r} 
\right)_{,pqij} + \epsilon_{ipq} \left(\frac{^{(2)} \mathcal S_{pm}}{r} 
\right)_{,qn}\epsilon_{jmn} \right. \nonumber\\
&&\left. + 2\left(\frac{^{(2)} \mathcal S_{p(i}}{r} \right)_{,j)p}
- \left(\frac{^{(4)} \mathcal S_{ij}}{r} \right)  \right]\;.
\label{framedragSpq}
\end{eqnarray}
Here $\mathcal S_{pq}$ is to be regarded as a function of retarded time, $t-r$, and the prefixes
$^{(2)}$ and $^{(4)}$ mean two time derivatives and four time derivatives.  This equation shows
explicitly how $\mathcal B_{ij}$ in the near zone transitions into $\mathcal B_{ij}$ in the
wave zone --- or equivalently, how rotating (or otherwise time-changing) frame-drag vortexes 
in the near zone generate gravitational waves.

This transition from near zone to far zone can also be described by the linear approximation
to the Maxwell-like equations for 
the frame-drag field $\bm{\mathcal B}$ and the tidal
field $\bm{\mathcal E}$, Eqs.\ (\ref{LLFMaxwell}). 
These equations govern the manner
by which the current-quadrupole near-zone frame-drag field (\ref{CurrentQuadrupoleB}) acquires
an accompanying tidal field as it reaches outward into and through the transition zone, to the
wave zone.  That accompanying tidal field is most easily deduced from the $\mathcal S_{pq}(t-r)/r$ terms in the metric perturbation, 
Eqs.\ (8.13) of \cite{thorne80}.  The result is:  
\begin{equation}
\mathcal E_{ij} = \frac 43 \epsilon_{pq(i}\left[-\left(\frac{^{(1)}
\mathcal S_{pk}}{r}\right)_{,j)kq}+\left(\frac{^{(3)}\mathcal S_{j)p}}{r}
\right)_{,q}\right]\;.
\label{tidalSpq}
\end{equation}

In the near zone, the current quadrupole's tidal field 
[first term of (\ref{tidalSpq})] behaves differently from its
frame-drag field [first term of (\ref{framedragSpq})]: it has 
one additional time derivative and one fewer
space derivative. As a result, \textit{the tidal field
is smaller than the frame-drag field in the near zone by a factor of
$r/\lambdabar$,} where $\lambdabar$ is the reduced wavelength of the emitted
gravitational waves.

As one moves outward through the near zone to 
the transition zone, where $r\sim\lambdabar$, the tidal field
increases in magnitude to become the same strength as the frame-drag field.
The frame-drag and tidal fields behave this way, because it
is the near-zone vortexes that generate the gravitational waves, as discussed
above.

In the wave zone, the general current-quadrupole (outgoing-wave) frame-drag field
(\ref{framedragSpq}) reduces to
\begin{equation}
\mathcal B_{\hat a\hat b} = \frac{4}{3r} \left[ ^{(4)}S_{\hat a\hat b}(t-r)
 \right]^{\rm TT}\;.
\label{BabWZSpq}
\end{equation}
Here the indices are confined to transverse directions (the surface of a sphere of
constant $r$) in the orthonormal basis $\bm e_{\hat \theta}$, 
$\bm e_{\hat\phi}$, and
``TT'' means ``take the transverse, traceless part''.  From the third of the Maxwell-like
equations (\ref{LLFMaxwell}), or equally well from the general current-quadrupole tidal
field, Eq.\ (\ref{tidalSpq}), we infer the wave-zone tidal field:
\begin{equation}
\mathcal E_{\hat a\hat b} =
\frac{4}{3r} \left[ \epsilon_{\hat c(\hat a} \, ^{(4)} \mathcal S_{\hat b)\hat c}
(t-r)\right]^{\rm TT}\;,
\label{EabWZSpq}
\end{equation}
where 
$\epsilon_{\hat c\hat d}$ is the 2-dimensional Levi-Civita tensor on the 
sphere. 
Since $\mathcal E_{\hat a\hat b} = \mathcal R_{\hat a \hat 0 \hat b \hat 0} = 
-\frac12 ^{(2)}h_{\hat a \hat b}^{\rm TT}$, where $h_{\hat a \hat b}^{\rm TT}$ is the
transverse, traceless gravitational-wave field, our wave-zone tidal distortion
(\ref{EabWZSpq}) agrees with the standard result for the wave-zone
current-quadrupole gravitational-wave
field (Eq.\ (4.8) of \cite{thorne80}).

\subsection{Rotating current quadrupole}
\label{sec:RotatingCurrentQuadrupole}

In this section, we will discuss the vortex and tendex lines of a 
rotating current quadrupole. 

A large rotating-current-quadrupole moment arises during the merger and
ringdown of the extreme-kick configuration of a binary black hole 
(a quasicircular binary made of identical black holes, whose 
spins are antialigned and lie in the orbital plane).
During the merger, the four vortexes associated with the initial holes' spins
get deposited onto the merged horizon's equator, and they then rotate 
around the final Kerr hole's spin axis at the same rate as their separation 
vector rotates, generating a large, rotating-current-quadrupole moment 
(paper III in this series).

As a simple linearized-gravity model of this late time behavior, 
imagine that at an initial time $t=0$, the two vortex-generating spins, of 
magnitude $S$, are separated by a distance $a$ along the $x$ axis and are 
pointing in the $\pm y$ direction --- i.e.\ they have the same configuration
as the static current quadrupole discussed in Sec.\ 
\ref{sec:StationaryCurrentQuadrupole} above. Then at
$t=0$, the spins' current-quadrupole moment has as its nonzero components 
$\mathcal S_{xy} = \mathcal S_{yx} = Sa$ [Eq.\ (\ref{Sxz}) with the spin
axes changed from $z$ to $y$]. As time passes, the spins' separation vector
and the spins' directions rotate at the same angular velocity $\omega$ so
the configuration rotates rigidly. Then it is
not hard to show that the current-quadrupole moment evolves as 
\begin{eqnarray}
\mathcal S_{xy} &=& \mathcal S_{yx} = Sa\cos(2\omega t)\,, \nonumber \\
\mathcal S_{xx} &=& - \mathcal S_{yy} = - Sa\sin(2\omega t)\,. 
\label{eq:SabRot}
\end{eqnarray}

It is straightforward to calculate the frame-drag field produced by this 
quadrupole moment using Eq.\ (\ref{framedragSpq}), and to then compute the
vortex lines and their vorticities.

\begin{figure}
\includegraphics[width=0.9\columnwidth]{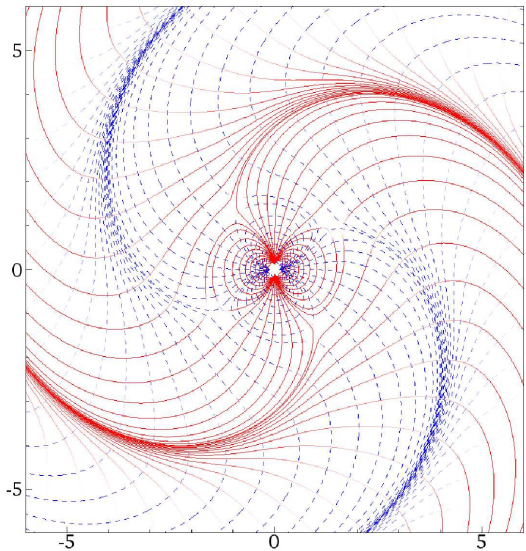}
\caption{
For a rotating current quadrupole in linearized theory, two families of
vortex lines in the
plane of reflection symmetry (the $x$-$y$ plane). 
The red (solid) curves are lines with negative vorticity, and the blue 
(dashed) curves are lines of positive vorticity. The color intensity 
of the curves represents the strength of the vorticity, but rescaled 
by $(k r)^5/[1 + (k r)^4]$ (with $k$ the wave number) to remove the
vorticity's radial decay. We see the quadrupolar near-zone pattern 
and the transition into the induction zone.  In the induction zone, 
the pattern carries four ``triradius'' singular points~\cite{Penrose1979} 
in each family of curves, necessitated for the transition from the 
static quadrupole pattern to the spiraling radiation pattern.
This same figure also describes the tendex lines of a rotating mass
quadrupole (see the end of Sec.~\ref{sec:MassQuadrupoleWaveGeneration}).
}
\label{fig:RotatingCurrentQuadrupoleVortex_near}
\end{figure}

The explicit expressions for these lines are somewhat lengthy, and not particularly instructive; but the shapes of the vortex lines and the values of their 
vorticities are quite interesting.   

\subsubsection{Vortex and tendex lines in the plane of reflection symmetry}

\begin{figure}
\includegraphics[width=0.9\columnwidth]{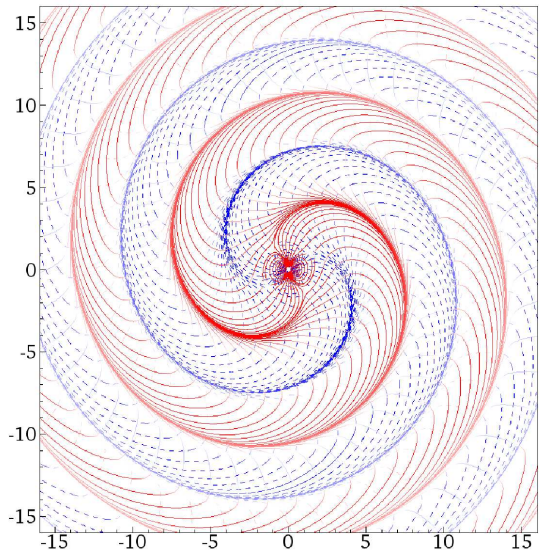}
\caption{
Same as Fig.\ \ref{fig:RotatingCurrentQuadrupoleVortex_near} but zoomed out
to show the wave zone. In the wave zone, the lines generically collect 
into spirals, which form the boundaries of vortexes (regions of concentrated vorticity).
}
\label{fig:RotatingCurrentQuadrupoleVortex_far}
\end{figure}

There are two sets of vortex lines that lie in the $x$-$y$ plane 
(the plane of reflection symmetry) and one set that passes orthogonally
through this plane.  We show the in-plane vortex lines in 
Figs.\ \ref{fig:RotatingCurrentQuadrupoleVortex_near}
and \ref{fig:RotatingCurrentQuadrupoleVortex_far}. 
The two figures depict the negative-vorticity vortex lines by red (solid) 
curves and the positive-vorticity lines by blue (dashed) curves.
The darkness of the lines is proportional to the vorticity; 
dark red (blue) indicates
strong negative (positive) vorticity, and
light red (blue) indicates weaker vorticity.
To remove the effects of the radial dependence in the coloring, we have scaled
the vorticity by $(kr)^5/[1+(kr)^4]$, 
where $k = 1/\lambdabar = 2\omega$ is the wave number of the
radiation.
Figure\ \ref{fig:RotatingCurrentQuadrupoleVortex_near} shows the region of the near zone 
that is difficult to see in Fig.\ \ref{fig:RotatingCurrentQuadrupoleVortex_far}, an
equivalent figure that spans a larger region of the $x$-$y$ plane.
As one can see from the figures, the two sets of lines have the same 
pattern, but are rotated with respect to each other by $\pi/2 = 90^{\rm o}$.

In the near zone (inner region of Fig.\ 
\ref{fig:RotatingCurrentQuadrupoleVortex_near}), 
the vortex-line pattern is the same as for the static current quadrupole
of Fig.\ \ref{fig:OmegafdSpq}b.
At the transition to the wave zone, 
the vortex lines fail to curve back into the central region and instead  
bend outward, joining a wave-zone spiral pattern.

That spiral pattern consists of four vortexes 
(regions of concentrated vorticity) that 
spiral outward and backward as the quadrupole rotates. These four regions 
of alternating positive and negative 
vorticity are bounded by tight clusters of vortex lines, just outside
of which the sign of the dominant vorticity changes. 

This same rotating vortex structure occurs in the case 
of an $l=2$, $m=2$, odd-parity (current-quadrupolar) perturbation
of a Schwarzschild black hole (Paper II in this series).
There the horizon 
vorticity pattern takes the place of the current quadrupole. 

\begin{figure}
\includegraphics[width=0.9\columnwidth]{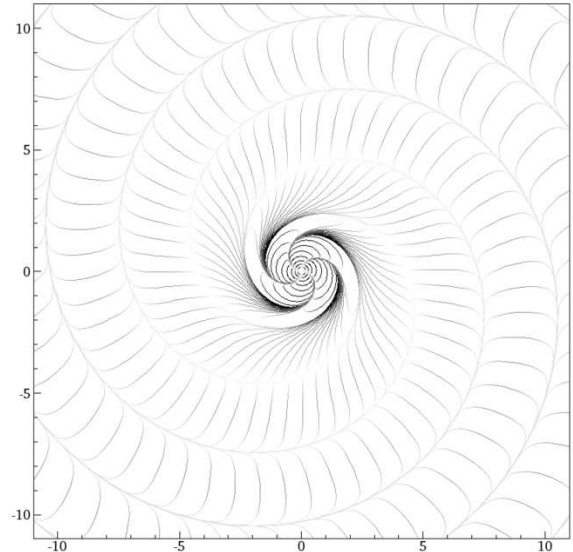}
\caption{
Tendex lines in the equatorial plane for a rotating current quadrupole 
in linearized theory.  The curves shown are lines of identically zero 
tendicity, enforced by symmetry.  The lines are shaded by the absolute value 
of the tendicity of the other two tendex lines that cross the lines shown, 
but are not tangent to the plane, and have equal and opposite tendicities.
}
\label{fig:RotatingCurrentQuadrupoleTendex_eq}
\end{figure}

In Fig.~\ref{fig:RotatingCurrentQuadrupoleTendex_eq} we indicate the 
structure of the tendex lines on the equatorial plane.  Because the 
symmetry properties of the system imply different constraints on the tendex 
field than on the vortex field, some explanation is needed.  The plane in 
which this and the previous two figures are drawn is a plane of reflection 
symmetry for the problem.  However, because the source is a pure current 
quadrupole, it must be antisymmetric under reflection across this plane (as 
such a reflection is a parity inversion).  The 
vorticity, which itself has an odd-parity relationship with its source, is 
symmetric under this reflection, constraining the vortex lines to be either 
tangent or orthogonal to the plane, as noted above.  The tendicity is 
antisymmetric under this reflection, so one family of lines can be tangent to 
the plane, so long as it has zero tendicity, and two other families of lines 
must cross the plane at equal and opposite inclinations, with equal and 
opposite tendicities, such that they are exchanged under the reflection.  The 
diagram in Fig.~\ref{fig:RotatingCurrentQuadrupoleTendex_eq} shows
the single family of tendex lines tangent to the symmetry plane.  As these 
curves have exactly zero tendicity, they are physically relevant only in that 
they denote the orientation of the other two families of tendex lines, which 
are not tangent to the plane, but whose projection onto the plane must be 
orthogonal to the curves shown (because all three curves are mutually 
orthogonal).  The shading of the lines in 
Fig.~\ref{fig:RotatingCurrentQuadrupoleTendex_eq} does not represent the 
tendicity of the lines drawn (which is identically zero), but rather of the 
other two tendex lines, which intersect the lines drawn with mutually 
equal and opposite tendicity.  Again, this shading is rescaled by 
$(kr)^5/[1+(kr)^4]$.  Though it is not apparent to the eye, the strength of 
the tendicity grows only as $r^4$ near the singular point (origin), rather than $r^5$ 
as for the vorticity.  As argued early in 
Sec.~\ref{sec:CurrentQuadrupoleWaveGeneration}, this can be interpreted 
intuitively as meaning that the vorticity is sourced directly from the 
current quadrupole, while the tendicity is sourced by induction from the 
time-varying vortex field.

For a rotating mass quadrupole (e.g.\ the quadrupole moment of an
equal-mass binary), the tendex lines in the plane of reflection
symmetry will have precisely the same form as the 
rotating-current-quadrupole vortex lines of Figs.\ \ref{fig:RotatingCurrentQuadrupoleVortex_near}
and \ref{fig:RotatingCurrentQuadrupoleVortex_far}; see Sec.\ 
\ref{sec:MassQuadrupoleWaveGeneration}.

\subsubsection{Vortex lines outside the plane of reflection symmetry:
Transition from near zone to wave zone}

\begin{figure}
\includegraphics[width=0.9\columnwidth]{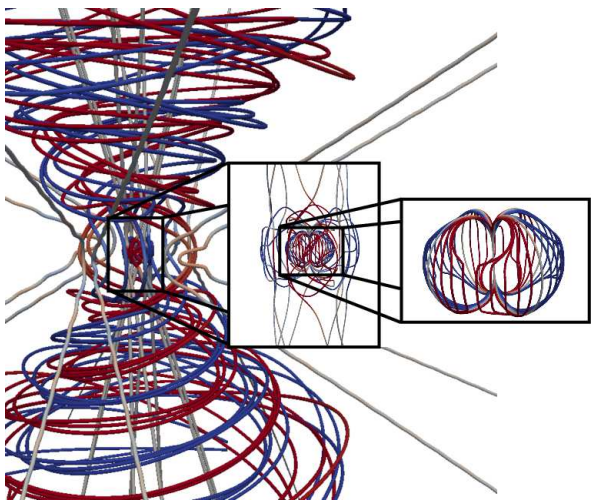}
\caption{
For the same rotating current quadrupole as in Figs.\
\ref{fig:RotatingCurrentQuadrupoleVortex_near} and 
\ref{fig:RotatingCurrentQuadrupoleVortex_far},
the family of vortex lines that pass orthogonally through the $x$-$y$
plane of reflection symmetry, color coded as in 
Fig.~\ref{fig:RotatingCurrentQuadrupoleVortex_near}.
In the wave zone, lines with approximately zero vorticity extend away from 
the source nearly radially, while lines with significant vorticity 
are dragged into tangled spirals by the rotation of the source.  In the left 
inset, we see the transition between the near and wave zones.  
Here, lines with nearly 
zero vorticity escape to infinity as in the wave zone, but those with 
significant vorticity are drawn toward the source.  The right inset delves 
down into the near zone, where the lines are approximately those 
of a stationary current quadrupole.  This same figure also describes the 
tendex lines of a rotating mass quadrupole (see the end of 
Sec.~\ref{sec:MassQuadrupoleWaveGeneration}).
}
\label{fig:RotatingCurrentQuadrupoleVortex_3d}
\end{figure}

Outside the plane of reflection symmetry and in the wave zone,  
the extrema of the vorticity show a spiraling pattern that is the same at
all polar angles.
More specifically, at all polar angles $\theta$, 
the magnitude of the vorticity, as a function of azimuthal angle $\phi$, has 
four maxima, and the locations of those maxima are the same as in the equator
($\theta=\pi /2$).
As in the equator, the maxima at fixed time $t$
spiral around at an angular rate
$d\phi_{\rm max}/dr = -\omega$ as one moves outward in radius, and as in the equator, 
vortex lines collect near these spiraling maxima,
and those lines too undergo spiraling behavior.

Figure \ref{fig:RotatingCurrentQuadrupoleVortex_3d} shows the development
of this spiraling structure as one moves outward from the near zone (innermost
inset) into the wave zone (outer region of figure).  This figure focuses on
the family of vortex lines that pass orthogonally through the
$x$-$y$ plane of reflection symmetry.
After entering the wave zone, the lines with nonnegligible vorticity 
(the blue and red lines) collect into a somewhat complicated spiral 
pattern, tangling among themselves a bit as they spiral. The 
gray lines with very low
vorticity, by contrast, point radially outward.  An animation of this rotating 
system can be seen at Ref.~\cite{RotatingCurrentQuadMovie}.  

It should be 
noted Fig.~\ref{fig:RotatingCurrentQuadrupoleVortex_3d}, and the animation 
at Ref.~\cite{RotatingCurrentQuadMovie}, represent 
somewhat incomplete descriptions of the structure of these field lines.  
The red and blue helical spirals shown in 
Fig.~\ref{fig:RotatingCurrentQuadrupoleVortex_3d} do not cross one another.  
However, at any point in space, there must be three mutually orthogonal vortex 
lines, with vorticities summing to zero.  Since at all points in the wave zone 
there is a field line of nearly zero vorticity directed in a nearly radial 
direction, through any point along these spirals of positive or negative 
vorticity, field lines of opposite vorticity must lie orthogonal to the spiral 
and to the approximately radial lines.  As shown in the following subsection, 
these lines form closed loops in the far-field region.

\subsubsection{Vortex lines in the far wave zone}

\begin{figure}
\includegraphics[width=0.9\columnwidth]{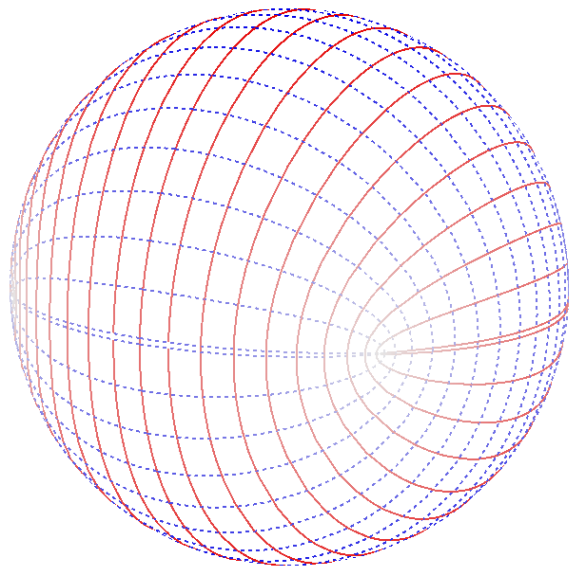}
\caption{(color online).
Vortex lines of a time-varying current quadrupole at very large $r$.
The lines are colored by the vorticity scaled by $r$, to remove the $1/r$
falloff, but the color coding is the same as in previous figures.
At very large distances from the source, the lines are transverse and
live on a sphere.
The third vortex line not shown is radial and has vanishing eigenvalue.}
\label{fig:CurrentQuadVortexInfinity}
\end{figure}

In the far wave zone (strictly speaking at future null infinity),
the frame-drag field becomes transverse
and traceless, and takes the simple form (\ref{BabWZSpq}).
Of its three sets of vortex lines, one is radial (with vanishing vorticity) and
the other two are tangent to a sphere of constant radius $r$
(with vorticity of equal and opposite
sign).
The two sets of vortex lines on the sphere have an interesting angular pattern
that is shown in Fig.\ \ref{fig:CurrentQuadVortexInfinity}.
The vortex line that lies in the equator alternates between positive
and negative vorticity, going to zero at four points (one of which is shown
at the front of the sphere).
This line is just the limit of the spirals where vortex lines collect
in Fig.\ \ref{fig:RotatingCurrentQuadrupoleVortex_far} at very large $r$. [Further discussion of the vortex and tendex lines of radiation at large $r$ is given in \cite{Zimmerman2011}, where the dual figure to Fig.~\ref{fig:CurrentQuadVortexInfinity} (tendex lines of a rotating mass quadrupole) is discussed in detail.]

\begin{figure}
\includegraphics[width=0.9\columnwidth]{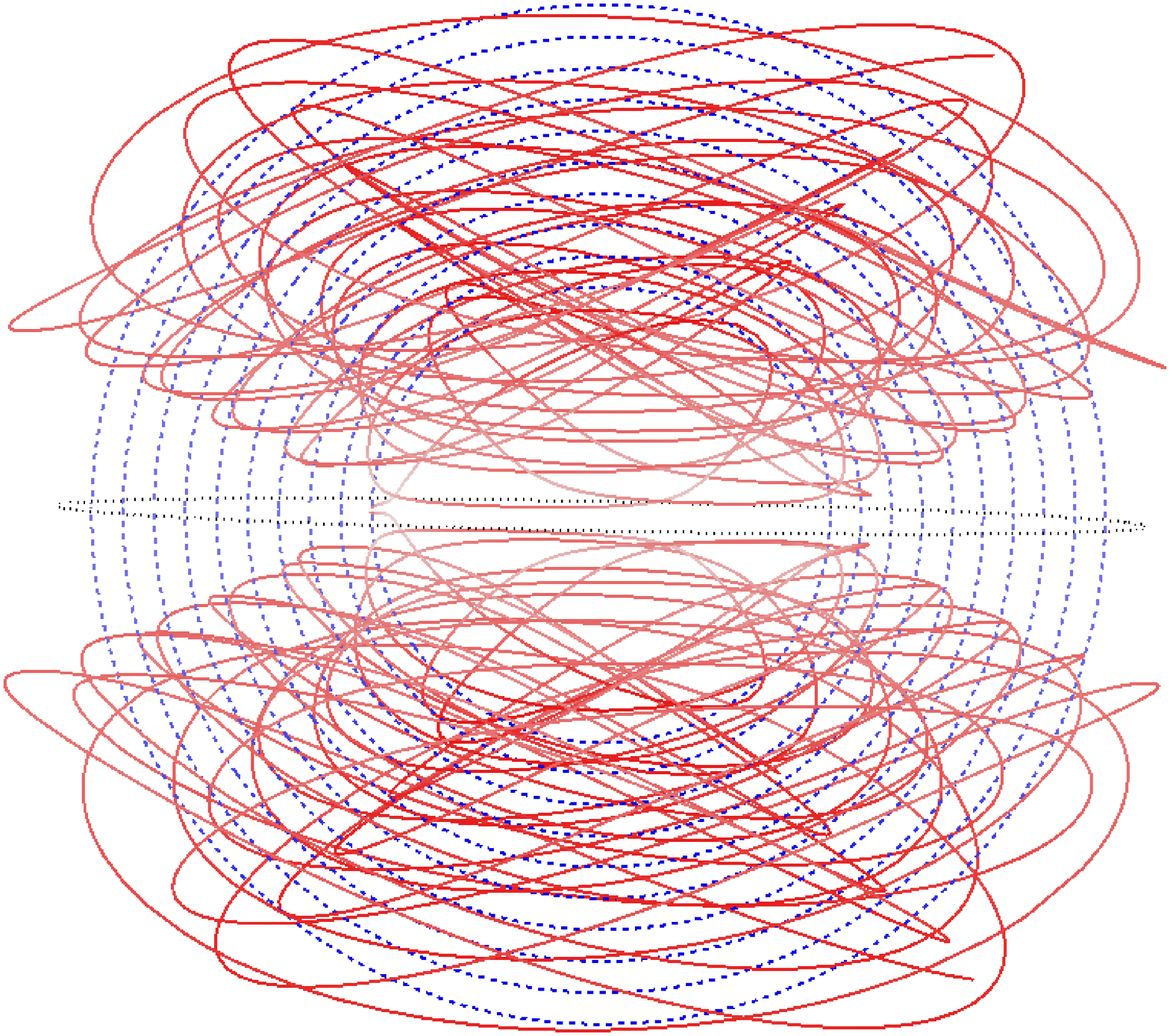}
\caption{(color online). 
Vortex lines of a rotating current quadrupole at sufficiently large
$r$ that the $1/r^2$ part of the frame-drag field may be thought of as a 
perturbation to the transverse vortex lines of Fig.\ 
\ref{fig:CurrentQuadVortexInfinity}.
The lines are colored by the vorticity as in that figure.
We also show a black dotted circle in the equatorial plane to identify
this plane.
The red solid lines shown here continue to collect on the 
maximum-vorticity spiral,
but they oscillate much more in polar angle than do the similar lines shown
in the near wave zone in Fig.\ \ref{fig:RotatingCurrentQuadrupoleVortex_3d}.
The blue dashed lines shown here form closed loops that pass from one 
positive-vorticity spiral to the next. 
This family of lines more closely resembles the transverse lines of Fig.\
\ref{fig:CurrentQuadVortexInfinity}, though in the limit of infinite radius,
the spiraling lines will also close to form transverse lines on the sphere.
There are also spiraling positive-vorticity (blue) lines and closed-loop,
negative-vorticity (red) lines, but to keep the figure from appearing muddled, 
we do not show them.
}
\label{fig:RotatingCurrentQuadVortex3D}
\end{figure}

How the vortex lines transition to the transverse pattern of Fig.\ 
\ref{fig:CurrentQuadVortexInfinity} at very large 
$r$, from the spiraling pattern of Fig.\ 
\ref{fig:RotatingCurrentQuadrupoleVortex_3d} in the inner wave zone,
is of considerable interest.  We can explore this by examining the
frame-drag field at sufficiently large radii that the
$1/r$ piece dominates over all other components, and that the 
$1/r^2$ part of the frame-drag field may be thought of 
as a perturbation to the leading-order $1/r$ part. In this region,
the vortex lines show two kinds of qualitative behavior.
Some of the vortex lines continue to form spirals that meander out and 
do not close, as in Fig.\ \ref{fig:RotatingCurrentQuadrupoleVortex_3d}.
There also are lines that form closed loops similar to the leading-order
vortex lines of Fig.\ \ref{fig:CurrentQuadVortexInfinity}.
We show both of these types of lines in Fig.\
\ref{fig:RotatingCurrentQuadVortex3D}.
The red, solid, spiraling lines continue to collect on the
maximum-vorticity spirals in the far wave zone.
These lines begin to resemble the transverse lines of Fig.\ 
\ref{fig:CurrentQuadVortexInfinity} more than the spiraling lines in the 
near wave zone of Fig.\ \ref{fig:RotatingCurrentQuadrupoleVortex_3d} do, 
because they rise and fall in polar angle as they wind around the 
maximum-vorticity spiral.
It is only in the limit of infinite radius that these spirals close to form
loops.
The blue, dashed, closed lines, on the other hand, resemble the
closed lines at infinity in Fig.\ \ref{fig:CurrentQuadVortexInfinity} much 
more closely.
The lines at finite $r$ do have some subtle differences between the 
corresponding lines at infinity: At finite radii, each individual line
passes from one maximum-vorticity spiral to the other; in doing so the line
must slightly increase in radius and rotate in azimuthal angle.
At the large radii shown in Fig.\ \ref{fig:RotatingCurrentQuadVortex3D},
this effect is very subtle.
We finally note that there are also spiraling, positive-vorticity lines and 
closed, negative-vorticity lines that we do not show to avoid visual clutter.

\subsection{Oscillating current quadrupole}
\label{sec:OscillatingCurrentQuadrupole}

The vortex lines of an oscillating current quadrupole (this section)
have a very different 
structure from those of the rotating current quadrupole (last section).
This should not be surprising, because the two quadrupoles arise
from very different physical scenarios, e.g., for the oscillating
quadrupole, the ringdown following a head-on collision of
black holes with antialigned spins, and for the rotating quadrupole,
the ringdown following the inspiral and merger of an extreme-kick 
black-hole binary. 
See Papers II and III of this series. 

In linearized theory, one can envision an oscillating current quadrupole as
produced by two particles, separated by a distance $a$ along the $x$ axis, 
whose spins, antialigned and pointing in the $\pm y$ direction, oscillate in
magnitude as $S\cos \omega t$.  The resulting quadrupole moment is [cf.\ 
Eq.\ (\ref{Spq})] 
\begin{equation}
\mathcal S_{xy} = \mathcal S_{yx} = Sa\cos\omega t\,.
\end{equation}

The frame-drag and tidal fields, and thence vortex and tendex lines, for this 
current quadrupole can be computed from Eqs.\ (\ref{framedragSpq}) and 
(\ref{tidalSpq}). 

\begin{figure}
\includegraphics[width=0.9\columnwidth]{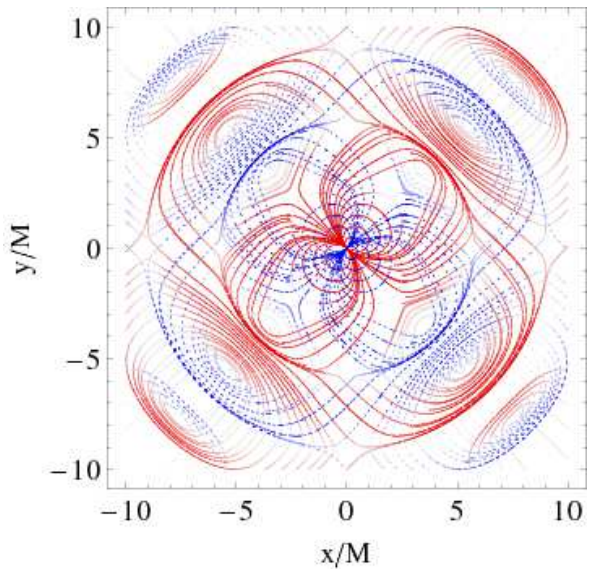}
\caption{(color online).
For an oscillating current quadrupole in linearized theory, two families
of vortex lines in the plane of reflection symmetry (the $x$-$y$ plane).
The color coding is the same as for the rotating current quadrupole,
Fig.~\ref{fig:RotatingCurrentQuadrupoleVortex_near}.
The vortex lines begin, near the
origin, like the static quadrupole pattern of Fig.~\ref{fig:OmegafdSpq}.
The effects of time retardation cause the pattern to stretch making larger
rectangular loops in the transition zone.
As time passes and the quadrupole oscillates, these loops detach from 
the origin and propagate out into the wave zone.
This same figure also describes the tendex lines of an oscillating mass
quadrupole (see the end of Sec.~\ref{sec:MassQuadrupoleWaveGeneration}).
}
\label{fig:OscillatingCurrentQuadNear}
\end{figure}

\begin{figure}
\includegraphics[width=0.9\columnwidth]{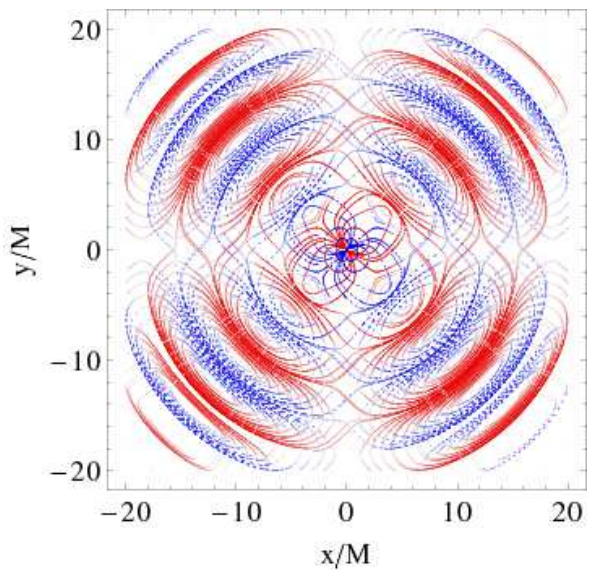}
\caption{(color online). Same as Fig.~\ref{fig:OscillatingCurrentQuadNear},
but zoomed out to show the wave zone.
Farther from the source, the loops take on a more regular alternating
pattern of gravitational waves.
The coloring shows that the vorticity is strongest at the 
fronts and backs of the loops, where the vortex lines are transverse to
the direction of propagation.
In the regions of the closed loops that extend radially, the field is weak 
(as one would expect for a transverse gravitational wave). 
}
\label{fig:OscillatingCurrentQuadFar}
\end{figure}

As for the rotating quadrupole, the $x$-$y$ plane of reflection symmetry
contains two families of vortex lines, and a third family passes orthogonally
through that plane. The in-plane vortex lines are depicted in Figs.\ \ref{fig:OscillatingCurrentQuadNear} and \ref{fig:OscillatingCurrentQuadFar} using
the same color conventions as for the rotating quadrupole (Figs.\ 
\ref{fig:RotatingCurrentQuadrupoleVortex_near} and
\ref{fig:RotatingCurrentQuadrupoleVortex_far}).
Figure \ref{fig:OscillatingCurrentQuadNear} shows the region of the near zone 
that is difficult to see in Fig.\ \ref{fig:OscillatingCurrentQuadFar}, an
equivalent figure that spans a larger region of the $x$-$y$ plane.
As one can see from the figures, the two families of vortex lines, solid red (negative
vorticity) and dashed blue (positive vorticity) have the same pattern, but are 
rotated by $\pi/2 = 90^{\rm o}$.

The way in which the gravitational waves are generated differs greatly from
the rotating current quadrupole of the previous section.
In the near zone, the two sets of vortex lines form a static quadrupole
pattern (identical to the near-zone rotating quadrupole of 
Fig.\ \ref{fig:RotatingCurrentQuadrupoleVortex_near},
but rotated by $\pi/4$
due to the orientation of the spins).
In the transition zone, 
the vortex lines form distorted loops that
head away from the origin, along the lines $y = \pm x$, in alternating 
fashion.
As they extend into the wave zone, 
the lines form two qualitatively different kinds of loops.
The majority of the loops reside only in one of the four quadrants of the
equatorial plane, but there are also loops that pass through all four
quadrants, staying near the regions of maximum vorticity, where lines collect
at the gravitational-wave crests.
For both types of loops, they maintain the same wavelength,
but the wave front becomes wider at larger radii, as they become
gravitational waves.
The portion of a loop transverse to the radial direction (the direction of
propagation) has strong vorticity, as one would expect for a gravitational
wave; in the radial portion of the loop, the vorticity is weak. Each cycle of the oscillating quadrupole casts off another set of vortex loops as the near-zone region passes through zero vorticity, and the loops travel outward towards infinity. 
This illustrates clearly the manner in which the near-zone vortex pattern generates gravitational waves in the far zone through its dynamics.

\begin{figure}
\includegraphics[width=0.95\columnwidth]{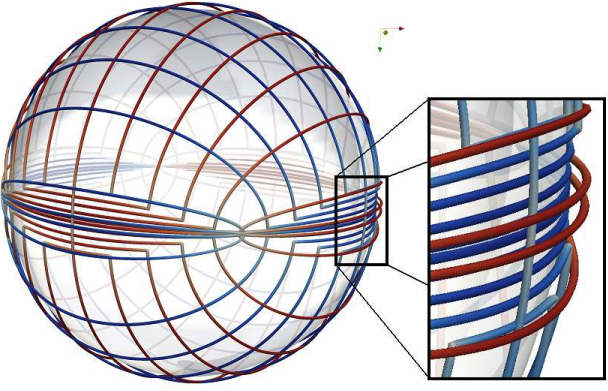}
\caption{(color online). 
Vortex lines of an oscillating current quadrupole at sufficiently large 
$r$ that the $1/r^2$ part of the frame-drag field may be thought of as a 
perturbation to the transverse vortex lines of Fig.\ 
\ref{fig:CurrentQuadVortexInfinity}.
The lines are colored in the same way as that figure, and the pattern of the 
lines around the poles is nearly identical to the transverse
lines of Fig.\ \ref{fig:CurrentQuadVortexInfinity}.
Near the equator, the $1/r^2$ perturbation causes the lines to bend and form
closed loops that reside in either the northern or the southern hemisphere.
The blue horizontal lines in the blow-up inset should be
compared with dense blue (dashed) bundles in 
Fig.~\ref{fig:OscillatingCurrentQuadFar}, and red lines  
with the red bundles immediately outside of the blue ones.
}
\label{fig:OscillatingCurrentQuad3D}
\end{figure}

As with the rotating current quadrupole, one can envision 
the equatorial vortex line of Fig.\ \ref{fig:CurrentQuadVortexInfinity} as 
the limit of the wave fronts of the planar vortex lines in Fig.\
\ref{fig:OscillatingCurrentQuadFar} at large distances.
It is again of interest to understand how the vortex lines outside the 
equatorial plane become the remaining vortex lines in Fig.\ 
\ref{fig:CurrentQuadVortexInfinity}.
To do so, we will make reference to Fig.\  \ref{fig:OscillatingCurrentQuad3D}, 
which shows the vortex lines at a distance sufficiently large that the 
$1/r^2$ portions of the frame-drag field can be thought of as a small 
perturbation to the transverse vortex lines of Fig.\ 
\ref{fig:CurrentQuadVortexInfinity}.
We show only the three-dimensional analog of the lines that pass
through all four quadrants in the equatorial plane, and do not show the
lines that remain in just one octant (analogous to the loops that remain
in one quadrant in the equatorial plane) to keep the figure as simple
as possible.

Near the poles, these vortex lines have nearly the same structure as the
purely transverse lines of Fig.\ \ref{fig:CurrentQuadVortexInfinity}; 
it is only near the equator that the lines begin to
differ.
As the lines approach the equator, 
they also increase in radius, due to the $1/r^2$ parts of the 
frame-drag field.
In doing so, they pass from one gravitational-wave crest to the next, 
and the lines sharply turn during their passage between successive crests.
The portion of the line on this next crest runs nearly parallel to
the equator, until it begins moving slightly inward (again due to the $1/r^2$
parts of the frame-drag field).
As it then sharply turns again, it returns to the original crest and begins 
heading back toward the poles.
This sharp turning happens on both sides of the sphere, which causes the lines
to form the closed loops that reside in either the northern or the southern 
hemisphere in Fig.\ \ref{fig:OscillatingCurrentQuad3D}.
Only in the limit that $r$ goes to infinity do the radial perturbations 
vanish, and the loops in the northern and southern hemisphere connect to
form the transverse pattern in Fig.\ \ref{fig:CurrentQuadVortexInfinity}.

\subsection{Wave generation by a time-varying mass quadrupole}
\label{sec:MassQuadrupoleWaveGeneration}

A time-varying mass-quadrupole moment $\mathcal I_{pq}(t)$ gives rise to 
metric perturbations of flat space given by the terms proportional to 
$\mathcal I_{pq}(t-r)/r$ and its derivatives in Eqs. (8.13) of \cite{thorne80}.
It is straightforward to calculate that the frame-drag field 
for these metric perturbations is 
\begin{equation}
\label{MagneticTensorMassQuad}
\mathcal B_{ij} = \epsilon_{pq(i}\left[\left(\frac{^{(1)}\mathcal I_{pk}}{r}
\right)_{,j)kq}-\left(\frac{^{(3)}\mathcal I_{j)p}}{r}\right)_{,q}\right]\;.
\end{equation}
Notice that this mass-quadrupolar frame-drag field is the same as the
current-quadrupolar tidal field  
(\ref{tidalSpq}), with the current-quadrupole moment 
$\mathcal S_{pq}$ replaced by $-\frac34
\mathcal I_{pq}$; cf.\ the duality relations (\ref{eq:DualityRotation}) and 
(\ref{duality}). 
Correspondingly, the vortex lines of this mass quadrupole will be the same as
the tendex lines of the equivalent current quadrupole.

The mass quadrupole's tidal field can be deduced from its frame-drag field
(\ref{MagneticTensorMassQuad}) by using the
third of the Maxwell-like equations
 (\ref{LLFMaxwell}). The result is 
\begin{eqnarray}
\label{ElectricTensorMassQuad}
\mathcal E_{ij} &=& \frac 12\left[-\left(\frac{\mathcal I_{pq}}{r} 
\right)_{,pqij} + \epsilon_{ipq} \left(\frac{^{(2)} \mathcal I_{pm}}{r} 
\right)_{,qn}\epsilon_{jmn} \right. \nonumber\\
&&\left. + 2\left(\frac{^{(2)} \mathcal I_{p(i}}{r} \right)_{,j)p}
- \left(\frac{^{(4)} \mathcal I_{ij}}{r} \right)  \right]\;.
\end{eqnarray}
Alternatively, this mass-quadrupolar tidal field can be deduced from the
current-quadrupolar frame-drag field (\ref{framedragSpq})
by using the duality relation
$\mathcal S_{pq} \rightarrow +\frac34 \mathcal I_{pq}$
[Eqs.\ (\ref{eq:DualityRotation}) and (\ref{duality})]. 

As a result, the tendex lines of this mass quadrupole will be the same as
the vortex lines of the current quadrupole, Figs.\ 
\ref{fig:RotatingCurrentQuadrupoleVortex_near} -
\ref{fig:RotatingCurrentQuadrupoleVortex_far} and 
\ref{fig:RotatingCurrentQuadrupoleVortex_3d}
- \ref{fig:OscillatingCurrentQuadFar}, with the red (solid) lines describing tidal stretching, and the blue 
(dashed) lines, tidal squeezing.

\subsection{Slow-motion binary system made of identical, nonspinning 
point particles}
\label{sec:Binary}

As a final example of a weakly gravitating system, 
we investigate the tendex lines of a Newtonian, equal-mass binary made
of nonspinning point particles in a circular orbit. 
We assume a separation $a$ between particles that is large compared to
their mass $M$, so the orbital velocity $v=\frac12 \sqrt{M/a}$ is small
compared to the speed of light (``slow-motion binary'').

Close to the binary, where retardation effects are negligible, the 
tidal field is given by the Newtonian expression $\mathcal E_{jk} = 
\Phi_{,jk}$ [Eq.\ \eqref{EijNewton}], with $\Phi$ the binary's Newtonian
gravitational potential
\begin{equation}
\label{BinaryStaticPot}
\Phi = - \frac{M_A}{|{\bm x} - {\bm x}_A|} -  \frac{M_B}{|{\bm x} - {\bm x}_B|} \,.
\end{equation}
Here $M_A=M_B = M/2$ are the particles' masses with $M$ the total mass,
and $\bm x_A$ and $\bm x_B$ are the locations of particles,
which we take to be on the $x$ axis, separated by a distance $a$. 

\begin{figure}
\includegraphics[width=0.95\columnwidth]{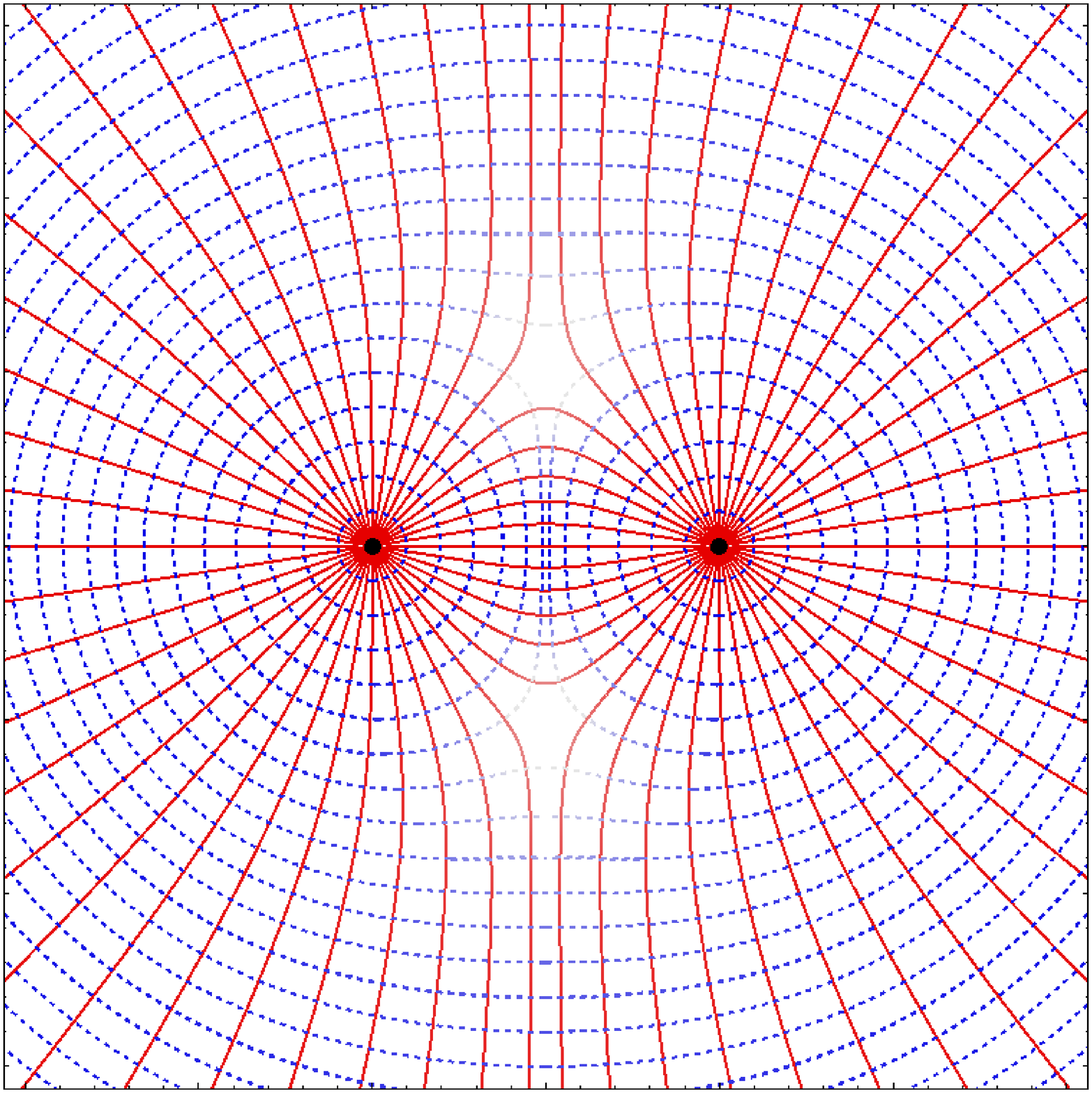}
\caption{(color online).
For a weak-gravity binary made of identical nonspinning point particles, in the 
near zone where retardation is negligible, two families of tendex lines
lying in a plane that passes through the two particles (e.g.\ the
orbital plane). The red (solid) curves are lines with negative tendicity,
and the blue (dashed) curves have positive tendicity.  The color intensity
of the curves represents the magnitude of the tendicity, rescaled by
$r_A^3r_B^3/[M^3(r_A^3 + r_B^3)]$, where $r_A$ and $r_B$ are the distances to 
the particles, to remove the tendicity's radial die out.
Near each particle, the tendex lines resemble 
those of an isolated spherical body; as one moves closer to the particle's
companion, the lines bend in response to its presence. At radii large compared
to the particles' separation $a$, the binary's monopole moment comes to 
dominate, and the tendex lines resemble those of a single isolated spherical
body.}
\label{fig:StaticBinary}
\end{figure}

In Fig.\ \ref{fig:StaticBinary}, we show
the near-zone tendex lines associated with this potential's tidal field,
color coded in the usual way (see the figure's caption). 
Close to each particle, the tendex lines resemble those of a static,
spherically symmetric object.
Moving farther from the particle, one can see the effects of the particle's companion,
bending and compressing the lines.  At radii $r \gtrsim a$, the Newtonian potential
and tidal field can be expanded in multipole moments with the monopole and
quadrupole dominating.  At $r>>a$, the monopole dominates and the tendex lines
become those of a single spherical body.

The binary's orbital angular velocity is $\omega = \sqrt{M/a^3}$
(Kepler's formula), and the binary emits gravitational waves with angular
frequency $2\omega$, reduced wavelength $\lambdabar = 1/(2\omega) = \frac12
\sqrt{a^3/M}$, and wavelength $\lambda=2\pi\lambdabar$.  
As a concrete example, we choose the particles' separation
to be $a=20M$; then $\lambdabar = \sqrt{5} a \simeq 2.24 a$, and 
$\lambda = 2\pi\sqrt5 a\simeq 14 a$.

\begin{figure}
\includegraphics[width=0.935\columnwidth]{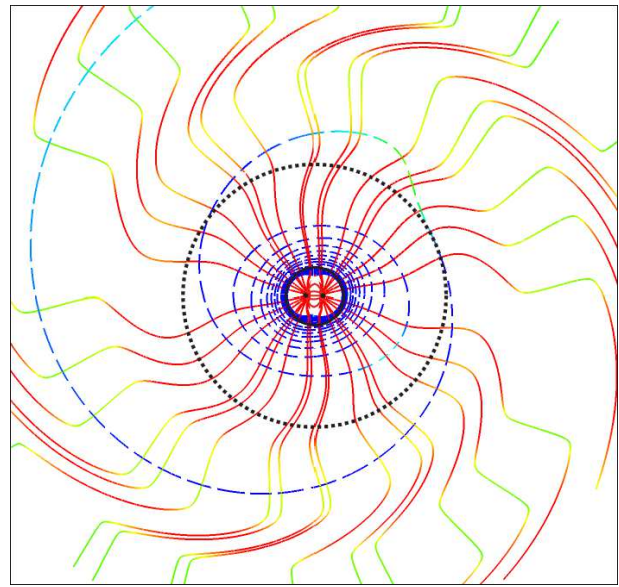}
\caption{(color online). 
Tendex lines in the orbital plane of the same binary as Fig.\ \ref{fig:StaticBinary}, with separation
$a=20M$ (where $M$ is the total mass), focusing on the
transition and wave zones $r\agt\lambdabar=2.24a$.  The solid black circle has
radius $\lambdabar$. The colors are fixed by the tendicity 
weighted by $\omega r$ so as to scale out the $1/r$ falloff in
the wave zone
(with dark blue 
strongly
positive, dark red strongly negative, and light green near zero). 
Inside the dotted black curve ($r=\frac12 a^2/M = 10a$), the binary's 
(nonradiative) monopole moment dominates, $\mathcal E \simeq M/r^3$ , and
the red (stretching) tendex lines are nearly radial.  Outside the dotted
black curve, the (radiative) quadrupole moment 
dominates, $\mathcal E \simeq 4M^3/a^4 r$, and the tendex lines are strong (significant tendicity) only
where they are approximately transverse to the radial direction.
}
\label{fig:BinaryWaveZone}
\end{figure}

Figure \ref{fig:BinaryWaveZone} shows tendex lines in this binary's orbital
plane, focusing on the transition and wave zones $r\agt \lambdabar =2.24a$ 
(outside the solid black circle). 
The shapes and colors of the tendex lines in this figure can be understood
in terms of the binary's multipole moments:

In the transition zone and wave zone, $r \agt \lambdabar$, the tidal field 
is the sum of a 
nonradiative monopolar piece with magnitude $\mathcal E_M \simeq M/r^3$,
and a quadrupolar piece with magnitude 
$(1/r) \partial^4\mathcal I/\partial t^4 \simeq (2\omega)^4 
(\frac14 M a^2)/r \simeq
4M^3/a^4 r $; higher order moments are negligible.  The two moments
contribute about equally at radius $r= \frac12 a^2/M = 10 a$ (dotted black
circle in the figure).  
The (nonradiative) monopole moment, with its red radial and blue circular 
tendex lines, dominates inside this circle. The (radiative) quadrupole 
moment dominates outside the circle, so there the tendicity is
significant (strong red and blue) only when the tendex lines are transverse;
strong red alternates, radially, with strong blue as the waves propagate
radially.  
Ultimately, at very large radii (far outside the domain of Fig.\ 
\ref{fig:BinaryWaveZone}), the quadrupole moment will totally dominate, 
and the tendex-line pattern will become that of a rotating quadrupole,
depicted in Fig.~\ref{fig:RotatingCurrentQuadrupoleVortex_far}.

\begin{figure}
\includegraphics[width=0.935\columnwidth]{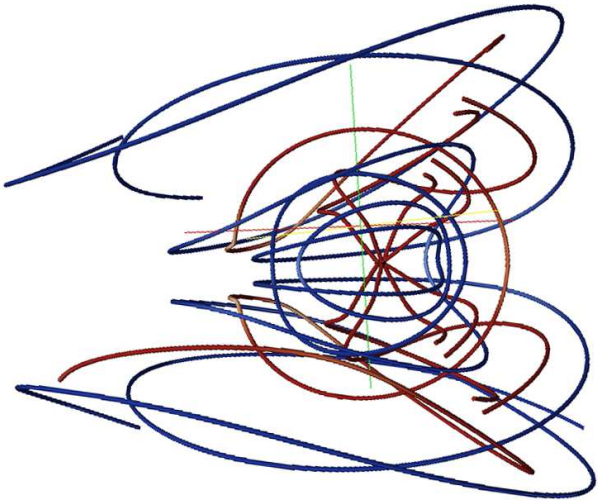}
\caption{(color online). 
Tendex lines outside the (central, horizontal) orbital plane, 
for the same binary and
parameters as Fig.\ \ref{fig:BinaryWaveZone}. 
In the inner region, the binary's 
monopole moment dominates, $\mathcal E \simeq M/r^3$, so
the red (stretching) tendex lines are nearly radial and the blue 
(squeezing) tendex lines are nearly circular.  At larger radii, 
the (radiative) quadrupole moment begins to be significant and then dominate,
so the tendex lines begin to spiral outward as for the rotating quadrupole
of Fig.\ \ref{fig:RotatingCurrentQuadrupoleVortex_3d}. 
}
\label{fig:BinaryWaveZoneOutOfPlane}
\end{figure}

Figure \ref{fig:BinaryWaveZoneOutOfPlane} shows the tendex lines for
this same binary, with the same parameters, in three dimensions, i.e.\ 
above and below the equatorial plane.  In the inner region, the monopole
moment dominates so the red (stretching) tendex lines are nearly radial,
and the blue (squeezing) tendex lines are nearly circular, centered on
the binary.  As one moves outward, the radiative quadrupole moment begins
to distort these radial and circular tendex lines, and then at large radii,
the now-dominant quadrupole moment drives them into the same spiraling
pattern as we have seen in Fig.\ \ref{fig:RotatingCurrentQuadrupoleVortex_3d}
for the tendex lines of a rotating, pure mass quadrupole.

\section{Conclusions}
\label{sec:Conclusions}

In this paper, we have focused on the electric and magnetic parts
of the Weyl curvature tensor, $\mathcal{E}_{ij}$ and $\mathcal{B}_{ij}$,
and have given them the names {\it tidal field} and {\it frame-drag field},
based on their roles in producing tidal gravitational accelerations
and differential frame dragging.  Being parts of the 
Riemann tensor, these fields
are well defined (though slicing dependent) in strong-gravity situations 
such as the near zone
of colliding black holes. For this reason, and because they
embody the full vacuum Riemann tensor and are easily visualized, 
$\mathcal E_{ij}$ and $\mathcal B_{ij}$ are 
powerful tools for exploring
the nonlinear dynamics of spacetime curvature (geometrodynamics). 

As tools for visualizing $\mathcal E_{ij}$ and $\mathcal B_{ij}$,
we have introduced
tendex and vortex lines (the integral curves of the
eigenvectors of $\mathcal{E}_{ij}$ and $\mathcal{B}_{ij}$), along with
their tendicities and vorticities (the eigenvectors' eigenvalues). 
The tendex and vortex lines are gravitational analogs of
electric and magnetic field
lines. Moreover, just as the electromagnetic field tensor
is fully determined by
its electric and magnetic field lines, together with
their field-line densities (which encode the lengths of the electric
and magnetic vectors), so the vacuum Riemann curvature tensor is fully
determined by its tendex and vortex lines, together with their colors
(which encode the tendicities and vorticities as in 
Fig.\ \ref{fig:BinaryWaveZone}).  

In terms of their transformation properties, the
($\mathcal{E}_{ij}$, $\mathcal{B}_{ij}$) pair is strictly analogous to
the pair of electric and magnetic 3-vector fields: they are components
of a 4-tensor, divided into two groups in a slicing dependent manner. We
are confident that this mild and transparent form of frame dependence
will not prevent our tendex and vortex concepts from becoming useful tools
for studying geometrodynamics,
any more than the frame dependence of electric and magnetic fields and 
field lines have been impeded these fields from being useful tools for 
studying electromagnetism in flat or curved spacetime. 

Using various examples from linearized gravity, for which analytical
formulas are available, we have plotted color-coded tendex and vortex
lines, and thereby we have gained insight into the behaviors of the tidal
and frame-drag fields. 
This intuition from
weak-gravity examples will be of great value when studying strongly 
gravitating
systems in asymptotically flat spacetimes, e.g.\ binary black holes.
This is because, 
in the weak-gravity region of spacetime outside such strong-gravity systems,
linearized gravity is a good approximation.  More specifically: 

For stationary, strongly gravitating systems (e.g., stationary black holes
and neutron stars), 
the tendex and vortex lines
in their asymptotic, weak-gravity regions 
will be well approximated by our linearized-theory results in 
Sec.~\ref{WeakGravityStationary}
(and, perhaps in some cases, extensions to higher multipoles). 

For oscillatory, strongly gravitating systems (e.g., binary black holes
and oscillating neutron stars), the wave zones' tendex and vortex lines
will be well approximated by those of our examples
in Sec.~\ref{sec:GWandGeneration}, and their extensions.  
Whether the
system has strong gravity or weak gravity, its wave-zone field lines
are controlled by radiative multipole moments that are tied to the
system's near-zone dynamics.  

As one moves inward through the weak-gravity wave zone into the
near zone and the region of strong gravity, the details of the field lines 
and the system's
dynamics may be quite different for strong-gravity systems than for our
weak-gravity examples. Nevertheless it seems likely that in {\it all} cases,
the gravitational waves will be generated by dynamical motions of 
near-zone tendexes
and vortexes (regions of strong tendicity and vorticity).  By exploring
that near-zone tendex/vortex dynamics, we can gain deep physical insight 
into nonlinear spacetime curvature. This will be a central theme of Papers
II and III in this series.

Whatever may be a source's strong-field dynamics, it will be useful
to focus on the imprints that the strong-field 
dynamics leaves on
the tendex/vortex structures in the strong-to-weak-gravity transition 
region.  Those transition-region tendex/vortex imprints will  
govern spacetime curvature throughout the asymptotic, weak-gravity region, 
and in particular will govern the radiative multipole moments that control the 
emitted
gravitational waves. 
Moreover, the imprinted structures in the strong-to-weak-gravity
transition region may turn out to have
some sort of effective dynamics that can be captured by simple analytical
models and can become a powerful tool for
generating approximate gravitational waveforms, e.g.\ for use in 
gravitational-wave data analysis.  

\acknowledgments
We thank John Belcher, Larry Kidder, Richard Price, and 
Saul Teukolsky for helpful discussions.  Our simulations have been
performed using the Spectral Einstein Code ({\tt SpEC})~\cite{SpECwebsite}.
This research was supported by
NSF grants PHY-0601459, PHY-0653653, PHY-0960291,
PHY-0969111, PHY-1005426, PHY-1068881 and CAREER grant PHY-0956189,
by NASA grants NNX09AF97G and NNX09AF96G,
and by  the Sherman Fairchild Foundation, the Brinson Foundation,
and the David and Barbara Groce fund.

\appendix

\section{The Newman-Penrose Formalism}
\label{sec:NPFormalism}

In this appendix we give the connection between the electric and magnetic
parts of the Weyl tensor $\bm {\mathcal E}$ and $\bm {\mathcal B}$, 
and the five Newman-Penrose (NP) curvature scalars~\cite{Newman1962}. 

The NP formalism~\cite{Newman1962} is especially useful 
for expressing the gravitational-wave content of a dynamical spacetime 
at asymptotic null infinity. It is also a crucial 
foundation for the study of black-hole perturbations and for the 
Petrov classification of vacuum spacetimes, both of which will naturally 
make contact with the study of vortexes and tendexes.  
In order to make contact with numerical simulations, we will need to 
understand the connection between the NP formalism and gravitational 
waves propagating on a flat background, as discussed in Sec.~\ref{sec:PlaneWave}. 

Because we use the opposite metric signature to that of the original 
Newman-Penrose paper~\cite{Newman1962} and the widely used Penrose-Rindler
book~\cite{Penrose1992}, our sign conventions for the NP quantities 
and for Eqs.\ (\ref{eq:Q}) and (\ref{eq:Psi2EB}) below differ
from theirs.  Ours are the same as in \cite{Stephani2003}. 

To begin with, we define an orthonormal tetrad 
$\vec{e}_{\hat \alpha}=(  \vec{e}_{\hat 0}, \vec {e}_{\hat 1}, 
\vec{e}_{\hat 2}, \vec{e}_{\hat 3})$ with time basis vector $\vec e_{\hat 0}
= \vec u$ orthogonal to our chosen foliation's spacelike hypersurfaces, and
with the spatial basis vectors $\vec {e}_{\hat 1}, \vec{e}_{\hat 2}, 
\vec{e}_{\hat 3}$ lying in those hypersurfaces. 
We use this tetrad to build a complex null tetrad for use in the NP formalism:
\begin{eqnarray}
\vec l = \frac{1}{\sqrt{2}}(\vec e_{\hat 0} + \vec e_{\hat 1}) \,, & &
\vec n = \frac{1}{\sqrt{2}}(\vec e_{\hat 0} - \vec e_{\hat 1}) \,, \nonumber \\
\vec m = \frac{1}{\sqrt{2}}(\vec e_{\hat 2} + i \vec e_{\hat 3})\,,  & &
\vec m^* = \frac{1}{\sqrt{2}}(\vec e_{\hat 2} - i \vec e_{\hat 3}) \,.
\label{NullTetrad}
\end{eqnarray}

By projecting the Weyl tensor onto this null basis, we construct the complex Weyl scalars,
\begin{subequations}
\begin{eqnarray} 
\label{eq:WeylScalar0}
\Psi_0 &=& C_{\mu \nu \rho \sigma} l^\mu m^\nu l^\rho m^\sigma \,, \\
\Psi_1 &=& C_{\mu \nu \rho \sigma} l^\mu n^\nu l^\rho m^\sigma \,, \\
\Psi_2 &=& C_{\mu \nu \rho \sigma} l^\mu m^\nu m^{* \rho} n^\sigma \,, \\
\Psi_3 &=& C_{\mu \nu \rho \sigma} l^\mu n^\nu m^{* \rho} n^\sigma \,, \\
\label{eq:WeylScalar4}
\Psi_4 &=& C_{\mu \nu \rho \sigma}n^\mu m^{* \nu} n^\rho m^{* \sigma} \,.
\end{eqnarray}
\end{subequations}
Using the null tetrad \eqref{NullTetrad} built from our orthonormal tetrad, 
we can express the spatial orthonormal components of the 
electric and magnetic parts of the Weyl tensor in terms of the 
Weyl scalars as follows:
\begin{eqnarray}
\label{eq:Q}
 &&\mathcal E_{\hat a \hat b} + i \mathcal B_{\hat a \hat b} \nonumber\\
  &=&  
\left[\begin{array}{ccc}
2 \Psi_2 & -(\Psi_1 - \Psi_3) &  i (\Psi_1 + \Psi_3) \\  
*  & \displaystyle \frac{\Psi_0+ \Psi_4 }{2}-\Psi_2 &\displaystyle  -\frac{i}{2}(\Psi_0-\Psi_4) \\
*  &*  &\displaystyle   - \frac{\Psi_0+\Psi_4}{2}-\Psi_2  \\
\end{array}
\right],
\end{eqnarray}
(cf.\ Eq (3.65) of \cite{Stephani2003}, where the differences are due to 
differing conventions on both $\bm{\mathcal B}$ and our null tetrad). 
In Eq.\ (\ref{eq:Q}), the rows and columns are ordered as 
$\hat 1, \hat 2, \hat 3$ and 
the entries indicated by $*$ are given by the symmetry of the matrix. 

The entries in Eq.\ (\ref{eq:Q}) can be derived in a straightforward manner 
from the definitions of $\bm{\mathcal E}$ and $\bm{\mathcal B}$, Eqs.~\eqref{eq:DefEij} and \eqref{eq:DefBij}, and the definitions of the Weyl scalars,  Eqs.~\eqref{eq:WeylScalar0}-\eqref{eq:WeylScalar4}. For example, we have
\begin{eqnarray}
\label{eq:EDerived}
\mathcal{E}_{\hat 1 \hat 1} & = & R_{\hat 1 \hat 0 \hat 1 \hat 0}
=\frac{1}{2}( R_{\hat r l \hat r l} + 2 R_{\hat r  l \hat r n} + R_{\hat r n \hat r n} )
\nonumber \\ & = &
\frac{1}{4}( R_{n l n l} - 2 R_{n l l n} +  R_{l n l n}) = R_{l n l n} \ ,
\end{eqnarray}
where we have used the symmetry properties of the Riemann tensor to eliminate and combine many terms. This result is not obviously equal to any of the Weyl scalars, but note that
\begin{eqnarray}
R_{l n l n} & = & - R^n{}_{n l n} = R^n{}_{n n l} = - (R^l{}_{n l l} +R^m{}_{n m l} + R^{m^*}{}_{n m^* l}) \nonumber \\
& = & - R_{m^* n m l} - R_{m n  m^* l} = R_{l m m^* n} + R_{l m^* m n} \nonumber \\
& = & \Psi_2 +  \Psi_2^* \,,
\end{eqnarray} 
where we have used the fact that in the null tetrad basis 
$\{\vec l, \vec n, \vec m, \vec m^*\}$, indices are raised and lowered with 
the metric components 
\begin{equation}
g_{\alpha \beta} = g^{\alpha\beta} =  
\left ( \begin{array}{cccc}
0 & -1 & 0 & 0 \\
-1 & 0 & 0 & 0 \\
0 & 0 & 0 & 1 \\
0 & 0 & 1 & 0 
\end{array} \right ) \,,
\end{equation}
and the fact that the Ricci tensor vanishes in vacuum spacetimes. 
Similar manipulations give
\begin{eqnarray}
\label{eq:BDer}
B_{\hat 1 \hat 1} & = & \frac 12 \epsilon_{\hat 1}{}^{\hat p \hat q} R_{\hat p \hat q \hat 1 \hat 0} = R_{\hat 2 \hat 3 \hat 1 \hat 0} =  - i R_{m^* m l n} \nonumber \\
&=&  i (R_{l m n m^*} + R_{l m^* m n}) = i (-\Psi_2 + \Psi_2^*) \,,  
\end{eqnarray}
so we see that $\mathcal{E}_{\hat 1 \hat 1} + i \mathcal{B}_{\hat 1 \hat 1} = 2 \Psi_2$. Similar computations give all of the entries of Eq.~\eqref{eq:Q}.

We will often have reason to consider the ``horizon tendicity'' and ``horizon vorticity.'' These are the values of $\bm{\mathcal E}$ and $\bm{\mathcal B}$ projected normal to the 2-dimensional event horizon of a spacetime containing a black hole, evaluated at the horizon. If the inward normal to the horizon is denoted $\bm N$ and we choose the vector ${\bm e}_{\hat 1}$ such that it coincides with $- \bm N$ at the horizon, then we immediately have the useful result
\begin{eqnarray}
\label{eq:Psi2EB}
\Psi_2 &=&  \frac12 ( \mathcal E_{NN} + i \mathcal B_{NN}) \nonumber \\
& = & \frac12  (\mathcal E_{ij} + i \mathcal B_{ij}) N^i N^j \,,
\end{eqnarray}
which we will use in our studies of analytic and numerical spacetimes 
containing horizons (papers II and III in this series). 

\bibliography{NicholsOwenZhangZimmermanEtAl.bbl}

\end{document}